\documentclass[11pt]{article}

\usepackage{cite}
 \usepackage{subfigure}
\usepackage{multirow}
\usepackage{helvet}
\usepackage{amsmath}
\usepackage{amssymb}
\usepackage{setspace}
\usepackage{graphicx}
\usepackage{empheq}
\usepackage{soul}
\usepackage{tabularx}
\usepackage{array}
\usepackage{float}
\usepackage{braket}
\usepackage[utf8]{inputenc}
\usepackage{url}
\usepackage{hyperref}
\usepackage{slashed}
\usepackage[font=footnotesize,labelfont=bf]{caption}
\usepackage{color}
\usepackage[dvipsnames]{xcolor}
\usepackage{soul}

\def\be{\begin{equation}}
\def\ee{\end{equation}}

\usepackage[a4paper,top=3cm,bottom=3cm,left=2.5cm,right=2.5cm,bindingoffset=0mm]{geometry}

\begin{document}

% Header ---------------------------------------------------
\hspace*{\fill} DESY-23-185

\begin{center}

{\Large \bf The Impact of LHC Jet and $Z$ $p_T$ Data at up to Approximate \\ \vspace{2mm} N${}^3$LO Order in the MSHT Global PDF Fit}

\vspace*{1cm}
T. Cridge$^a$, L. A. Harland-Lang$^{b}$, 
and R. S. Thorne$^b$\\                                               
\vspace*{0.5cm}                                                    
   
$^a$ Deutsches Elektronen-Synchrotron DESY, Notkestr. 85, Hamburg 22607, Germany \\
$^b$ Department of Physics and Astronomy, University College London, London, WC1E 6BT, UK     

\begin{abstract}
\noindent  We present an analysis of two key sets of data constraining the high $x$ gluon at up to approximate N${}^3$LO  in QCD within the MSHT global PDF fitting framework. We begin with  LHC 7 and 8 TeV inclusive jet and dijet production at both NNLO and aN${}^3$LO. This makes use of the formalism established in the previous global MSHT20aN${}^3$LO PDF fit, but now considers the role of dijet production for the first time at this order.  We present a detailed comparison of the fit quality and PDF impact for both cases, and consider the role that electroweak corrections, and the scale choice for inclusive jet production has. Some mild tension between these data sets in the impact on the high $x$ gluon is seen at NNLO, but this is largely eliminated at aN${}^3$LO. While a good fit quality to the dijet data is achieved at both orders, the fit quality to the inclusive jet data is relatively poor. We examine the impact of including full colour corrections in a global PDF fit for the first time, finding this to be relatively mild. We also revisit the fit to the ATLAS 8 TeV $Z$ $p_T$ data, considering the role that the $p_T$ cuts, data selection and different aspects of the aN${}^3$LO treatment have on the fit quality and PDF impact.  We observe that in all cases the aN${}^3$LO fit quality is consistently  improved relative to the NNLO, indicating a clear preference for higher order theory for these data.

\end{abstract}

\end{center}

\section{Introduction}

Parton distribution functions (PDFs) of the proton are a key element in the LHC precision physics program. As such, recent dedicated efforts to extract these as accurately and precisely as possible have been performed by multiple groups, accompanied by the release of public PDF sets~\cite{Bailey:2020ooq,NNPDF:2021njg,Hou:2019efy,ATLAS:2021vod}.  In these analyses, a wide range of data from HERA and fixed target experiments to the Tevatron and the LHC are included in the fit, while the state--of--the--art in the theoretical calculation entering these fits is now by default next--to--next--to leading order (NNLO) in the QCD perturbative expansion.

More recently the first PDF analysis at approximate ${\rm N}^3$LO order was performed by the MSHT group~\cite{McGowan:2022nag} and publicly released in the \texttt{MSHT20an3lo} PDF set. This accounted for the significant amount of known information about the ${\rm N}^3$LO results for the PDF evolution, heavy flavour transitions and DIS coefficient functions, while also including approximations for the unknown parts, with corresponding uncertainties associated with these. A clear impact on the fit quality was observed, with as expected an improvement seen by going to this next (approximate) order, while the impact on the PDFs and LHC phenomenology was in some cases found to be significant. Given the amount of known ${\rm N}^3$LO information available, these represent a more accurate PDF determination. An analysis of the impact of QED corrections on top of the a${\rm N}^3$LO fit was also presented recently \cite{Cridge:2023ryv}, though these are not included in this work.

There are many implications for LHC physics and PDF fitting to investigate. A particularly relevant topic relates to the impact of LHC data on the high $x$ gluon. This is a region where direct information was relatively lacking prior to the inclusion of data from hadronic collisions, and specifically the LHC, in PDF fits. However, with the advent of the LHC and the availability of a range of differential and high precision inputs from, in particular, top quark pair, jet and $Z$ boson $p_T$ (or $Z$ boson in association with jets) measurements, significant constraints on the high $x$ gluon, as well as other parton flavours, can be placed. The impact of theses data sets and their interplay has been the focus of much discussion, see~\cite{Harland-Lang:2017ytb,Bailey:2019yze,Bailey:2020ooq,Thorne:2022abv} for studies  in the context of the MSHT PDF fit, as well as elsewhere in~\cite{Hou:2019efy,Jing:2023isu,NNPDF:2017mvq,NNPDF:2021njg,PDF4LHCWorkingGroup:2022cjn,Cridge:2021qjj,Amoroso:2022eow}. 

In the case of jet production, only inclusive jet data are included in the baseline MSHT20 fit~\cite{Bailey:2020ooq}, as well as the more recent a${\rm N}^3$LO analysis~\cite{McGowan:2022nag}. However, a range of LHC dijet data are available~\cite{ATLAS:2013jmu,CMS:2012ftr,CMS:2017jfq,ATLAS:2017ble,CMS:2022wdi} along with NNLO theoretical calculations to match them~\cite{Currie:2016bfm,Gehrmann-DeRidder:2019ibf}.  Indeed, the impact of dijet data on the NNPDF fits has been studied in~\cite{AbdulKhalek:2020jut,NNPDF:2021njg} at up to NNLO. While in~\cite{AbdulKhalek:2020jut} full consistency was found between the inclusive jet and dijet data sets, in~\cite{NNPDF:2021njg} the CMS 8 TeV triple differential data~\cite{CMS:2017jfq}  was found to be in significant tension with the baseline fit excluding it, in particular in terms of the impact on the high $x$ gluon. This is of particular note as these data are the only case so far to be included in a PDF fit that are triple differential in the dijet kinematics; this provides a greater constraining power on the PDFs. 

In light of this, it is highly relevant to examine the impact of these dijet data on the MSHT20 fit. As well as providing an assessment of this impact within the context of a distinct PDF fit, a key novel element we will examine here is the effect that going to approximate ${\rm N}^3$LO will have. By doing so, we can determine to what extent questions of consistency between the jet and dijet case, and between these and other data entering the fit, as well as their PDF impact, changes in going to the more accurate a${\rm N}^3$LO order. In the latter case, theoretical uncertainties from  missing higher orders in the calculation of the cross sections are also, crucially, included. Indeed, as discussed in~\cite{McGowan:2022nag,Jing:2023isu}, the effect of going to a${\rm N}^3$LO can be significant in terms of the PDF impact of given data sets and the tensions between them.

A further relatively recent theoretical advance that it is interesting to consider is the impact of including full colour corrections at NNLO, as first calculated in~\cite{Chen:2022tpk}. These are shown in this reference to be moderate but not negligible in particular for the case of triple--differential dijet production. However, thus far they have not been considered within the context of a global PDF fit. We therefore study the impact of these corrections here.

In addition to jet and dijet data, as discussed above a further data set of relevance to the high $x$ gluon is the ATLAS 8 TeV $Z$ $p_T$ data~\cite{ATLAS:2015iiu}. This was observed in~\cite{Bailey:2020ooq} to be in some tension with other data sets in terms of its impact on the gluon, most notably with differential top quark pair production and inclusive jet data.  The fit quality was found to be rather poor at NNLO, but in~\cite{McGowan:2022nag} this was observed to improve dramatically when going to a${\rm N}^3$LO. In addition, it was illustrated in \cite{Jing:2023isu} that the tension between this data set and others in the fit seemed to reduce at a${\rm N}^3$LO. 
At NNLO, the fit quality and impact of this data set were found in the CT and NNPDF fits~\cite{Hou:2019efy,NNPDF:2021njg} to be somewhat different, however here smaller subsets of the data were fit to, as well as there being other differences in the treatment of the theory (in particular the uncertainties in the NNLO K-factors) entering the fit, see \cite{Bailey:2020ooq} for more details. Given this, a more detailed analysis of this data set, and its impact on the MSHT fit, is again  well motivated.

In light of the above discussion, in this paper we present the first analysis of inclusive jet and dijet production at up to a${\rm N}^3$LO order. We analyse in detail the fit quality, consistency between the jet and dijet cases, and overall PDF impact at up to a${\rm N}^3$LO order. We also study the effect of including electroweak (EW) corrections, the choice of scale in the case of inclusive jet production (namely $p_\perp^{\rm jet}$ of $H_\perp$), and the impact of full colour corrections in the case of dijet production. We in addition investigate the effect of how modifying the manner in which the ATLAS 8 TeV $Z$ $p_T$ data~\cite{ATLAS:2015iiu} is treated affects the fit at up to a${\rm N}^3$LO order, namely by  increasing the lower cut on  $p_T^{ll}$ in order to assess any potential limitations of the NNLO fit to these data, as well providing a rather closer comparison to the data subsets that are fit by the CT and NNPDF groups. We also consider the extent to which the known aN${}^3$LO information contributes to the improved fit.

The outline of this paper is as follows. In Section~\ref{sec:datatheoryjet} we present the inclusive jet and dijet data sets that enter the fit, as well as the theoretical calculation corresponding to these. In Section~\ref{sec:fitquality} we discuss the fit quality at up to a${\rm N}^3$LO order for both inclusive jet and dijet cases. In Section~\ref{sec:PDFs} we present their impact on the high $x$ gluon. In Section~\ref{sec:zptdatatheory} we discuss the $Z$ $p_T$ data and the corresponding theory in our study. In Section~\ref{sec:zptfitquality} we present the relevant fit qualities and in Section~\ref{sec:zptPDFs} the PDFs that result from this. Finally, in Section~\ref{sec:conc} we conclude.

\section{Jet and Dijet production at NNLO and aN${}^3$LO}

\subsection{Data and Theory}\label{sec:datatheoryjet}

The MSHT20 fit~\cite{Bailey:2020ooq} includes inclusive jet data from the Tevatron~\cite{D0:2011jpq,CDF:2007bvv} and CMS at 2.76 TeV~\cite{CMS:2015jdlc}, as well as data from ATLAS~\cite{ATLAS:2014riz} at 7 TeV and CMS~\cite{CMS:2014nvq,CMS:2016lna} at 7 and 8 TeV. In all fits which follow, we continue to include the Tevatron and CMS 2.76 TeV data (for which there are no dijet data counterparts), as well as all other non--jet data sets in the MSHT20 fit. We will supplement this by either the 7 and 8~TeV LHC inclusive jet data or their counterpart dijet data, which cannot be included simultaneously due to their statistical overlap, as discussed later.

For the inclusive jet fits, we include the ATLAS 7 TeV and CMS 7 and 8 TeV jet data that are in the MSHT20 fit, and now include also the ATLAS 8 TeV~\cite{ATLAS:2017kux} data set. We take the larger of the jet radii available, namely $R=0.6$ for the ATLAS data, and $R=0.7$ for the CMS. We choose the larger jet radius as this seems to be slightly more perturbatively convergent, although hadronic corrections are a little larger, see e.g.~\cite{ATLAS:2017kux}. For the ATLAS 7 and 8 TeV data we apply a smooth decorrelation of certain systematic error sources, guided by the proposal described in~\cite{ATLAS:2017kux}. The implementation of this is exactly as described in~\cite{Bailey:2020ooq} for the 7 TeV data, while for the 8 TeV data we use this form of decorrelation, but applied to four sources, as suggested in~\cite{ATLAS:2017kux}, namely the jet flavour response, the multi--jet balance fragmentation, the jet energy scale pile--up $\rho$ topology, and the non--perturbative corrections.

\begin{figure}[h]
\begin{center}
\includegraphics[scale=0.6]{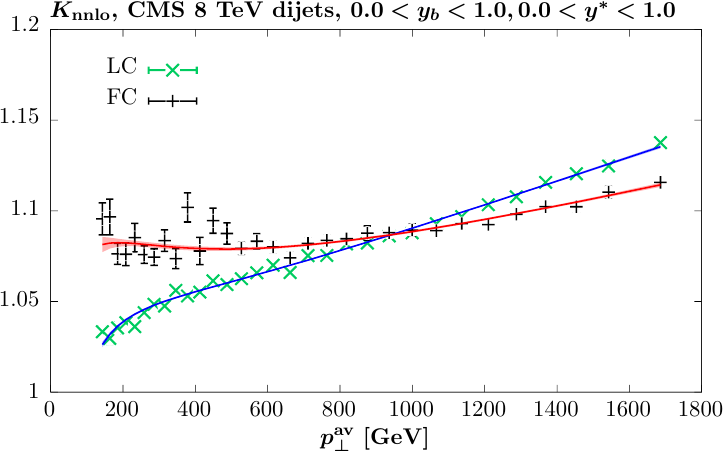}
\includegraphics[scale=0.6]{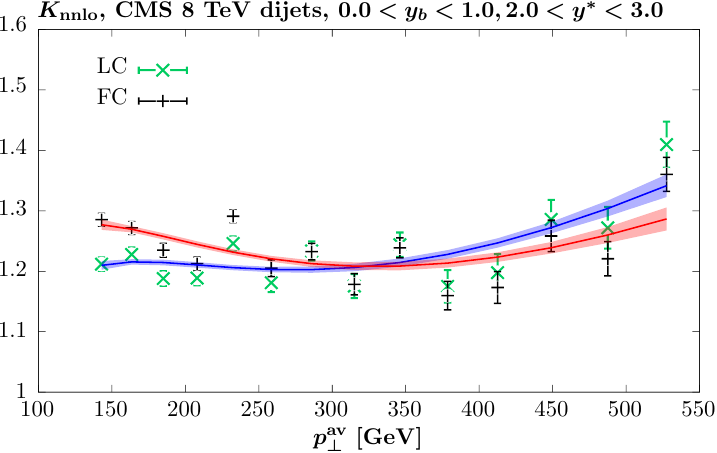}
\includegraphics[scale=0.6]{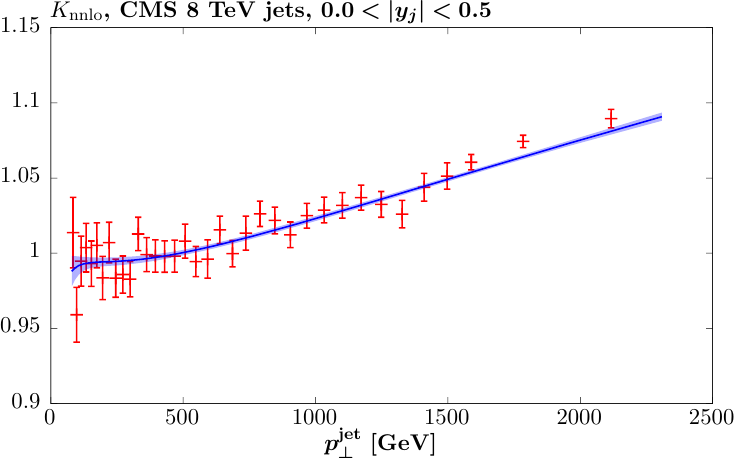}
\includegraphics[scale=0.6]{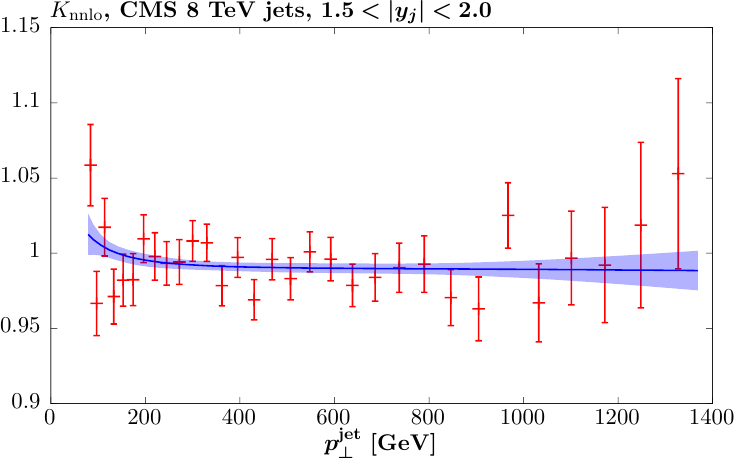}
\caption{\sf The calculated NNLO to NLO K-factors, including the MC statistical errors,
corresponding to a selection of the CMS 8 TeV inclusive jet~\cite{CMS:2016lna} and dijet~\cite{CMS:2017jfq} data, in the latter case at leading (LC) and full (FC) colour in the NNLO calculation. Also shown is the 4–parameter
fit to the K-factors, following the treatment in~\cite{Harland-Lang:2017ytb}. The uncertainty band is shown for illustration by
summing in quadrature the 68\% C.L. fit uncertainties in each bin, i.e. omitting correlations. In the upper plots, the blue (red) bands correspond to the LC (FC) fits.}
\label{fig:kfs}
\end{center}
\end{figure}

The CMS 7 TeV data correspond to an updated analysis of the earlier 7 TeV data presented in~\cite{CMS:2012ftr}. For this earlier analysis the $p_\perp$ threshold is higher, with $p_\perp^j > 114$ GeV, while for the more recent analysis~\cite{CMS:2014nvq} this extends down to $p_\perp^j > 56$ GeV. The $p_\perp^j$ binning in the overlapping region is identical in the two cases. For the CMS 8 TeV data~\cite{CMS:2016lna} we include the full statistical correlations between bins, available on the \texttt{xFitter} website~\cite{xfitterweb}, following advice from the relevant analysers~\cite{Engin}. In all cases, the data are presented double differentially in the jet $p_\perp$ and rapidity, $y$, and reconstructed using the anti--$k_\perp$ algorithm~\cite{Cacciari:2008gp}, while we apply non--perturbative corrections and uncertainties provided by the experimental collaborations.

For the dijet fits, we include ATLAS data~\cite{ATLAS:2013jmu} at 7 TeV, and CMS data at 7 and 8 TeV~\cite{CMS:2012ftr,CMS:2017jfq}. The ATLAS data are presented double differentially in terms of the dijet invariant mass, $m_{jj}$, and half rapidity separation, $y^* = |y_1-y_2|/2$, of the two leading jets. The CMS 7 TeV data 
are presented double differentially in terms of the dijet invariant mass, $m_{jj}$, and maximum absolute rapidity, $|y_{\rm max}|$, of the two leading jets. The CMS 8 TeV data are presented triple differentially, in terms of the average transverse momentum, $p_{\perp {\rm avg}}$, half rapidity separation, $y^*$, and boost of the dijet system, $y_b=|y_1 + y_2|/2$, defined in all cases with respect to the leading jets. As we will see, the triple differential nature of the CMS 8 TeV data leads to this set having a particularly significant impact on the fit in comparison to the other dijet sets, due to its ability to isolate specific regions of parton $x$ more precisely. In the  ATLAS case we take the $R=0.6$ data set, while the CMS data use jet radius $R=0.7$. In all case the jets are again  reconstructed using the anti--$k_\perp$ algorithm~\cite{Cacciari:2008gp}, while we apply non--perturbative corrections provided by the experimental collaborations. For the CMS 8 TeV data we note that, as demonstrated in Appendix A.3 of~\cite{Sieber:2016kri}, a selection of the correlated systematic errors change sign at a given point in $p_\perp^{\rm av}$, however this is not reflected in the corresponding Hepdata entry. Following discussion with the relevant analysers~\cite{Klaus}, we have corrected these in our fit. As we will see, the impact on the fit quality is relatively minimal, which can be expected as the majority of these error sources change sign at larger $p_\perp^{\rm av}$, where the data are less precise.

We note that the inclusive jet fit does not include these dijet data sets, and likewise the dijet fits exclude the ATLAS and CMS 7 and 8 TeV inclusive jet data sets. This is necessary in the case of the CMS inclusive jet data, and the ATLAS 7 TeV inclusive jet data, in order to avoid double counting, given the correlations between the inclusive and dijet measurements are currently not available, and these do not come from distinct underlying data sets. In principle, this is not the case for the ATLAS 8 TeV data, where no dijet measurement is available at the time of this study, but in order to better differentiate between the fitting of jet and dijet data, we continue to exclude this in the dijet fit.

For the theoretical predictions we use NLO \texttt{APPLGrid}~\cite{Carli:2010rw} and \texttt{FastNLO}~\cite{Kluge:2006xs,Britzger:2012bs} grids, supplemented by NNLO K--factors calculated using \texttt{NNLOJET}~\cite{Currie:2016bfm,Gehrmann-DeRidder:2019ibf}. For the NNLO corrections, the leading colour (LC) results are used in all cases except for the CMS 8 TeV dijet data~\cite{CMS:2017jfq}, where the full colour (FC) predictions~\cite{Chen:2022tpk} are also compared to. As discussed in~\cite{Chen:2022tpk} the impact of FC at NNLO is small in the case of both inclusive jet and double differential dijet production, therefore we only expect any significant effect to occur for this triple differential data set. Moreover for many of the other cases the FC results are not publicly available. Given this, we will take the LC results as our baseline in the dijet fit, although we will examine all relevant differences that come from instead using the FC result in the dijet fit.

\begin{figure}[t]
\begin{center}
\includegraphics[scale=0.6]{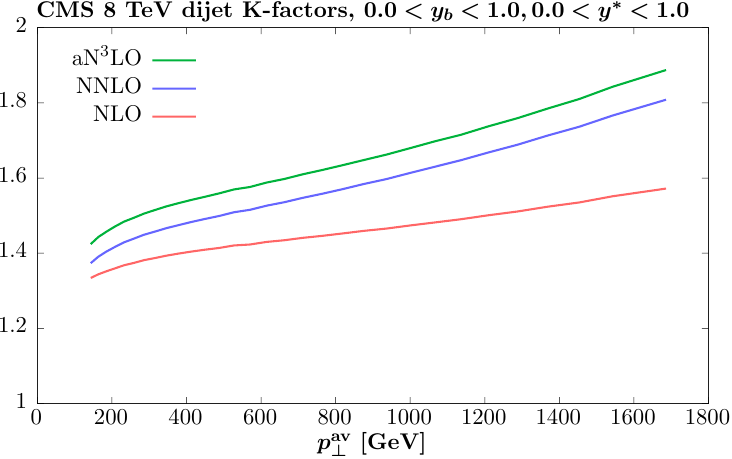}
\includegraphics[scale=0.6]{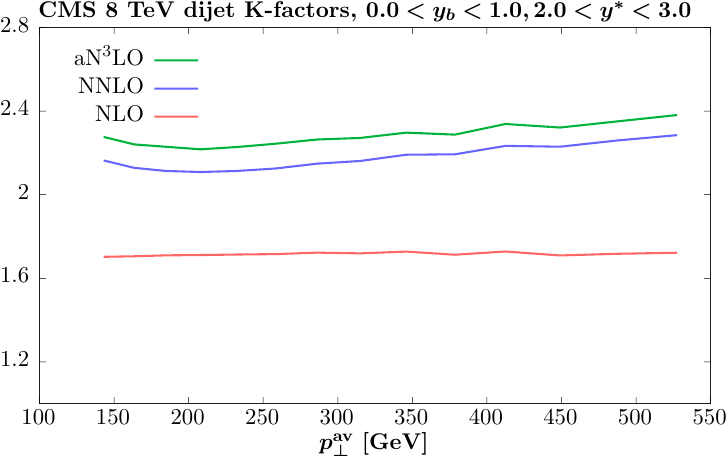}
\caption{\sf The NLO, NNLO and aN${}^3$LO QCD K-factors corresponding to a selection of the CMS 8 TeV dijet~\cite{CMS:2017jfq} data. The aN${}^3$LO QCD K-factors correspond to the default dijet fit described in Section~\ref{sec:fitquality}, and the NNLO corrections to the LC result.}
\label{fig:kn3lo}
\end{center}
\end{figure}

\begin{figure}[t]
\begin{center}
\includegraphics[scale=0.6]{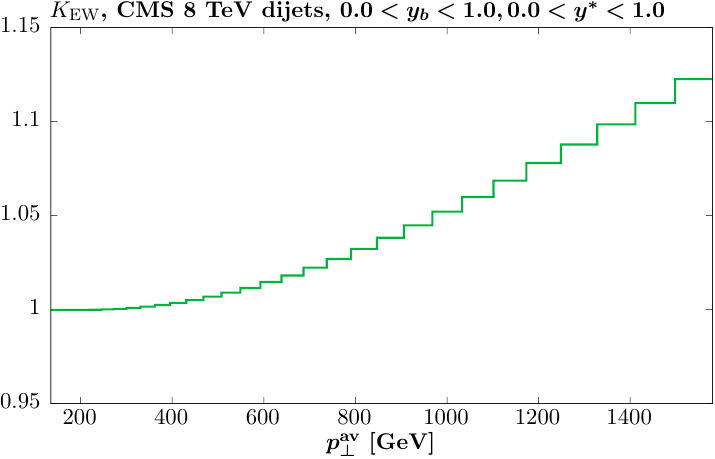}
\includegraphics[scale=0.6]{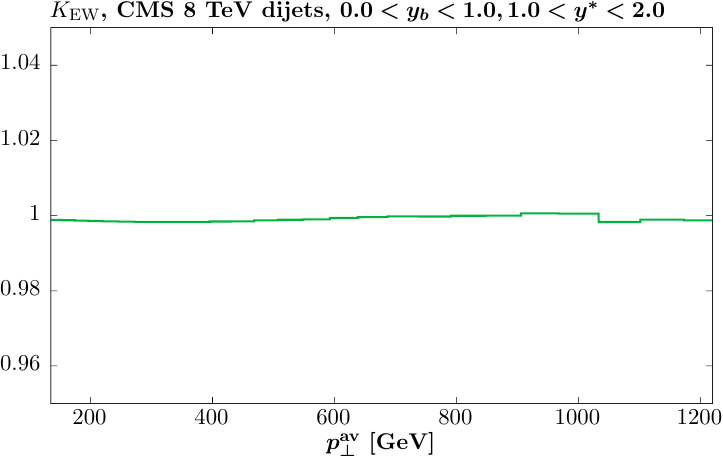}
\includegraphics[scale=0.6]{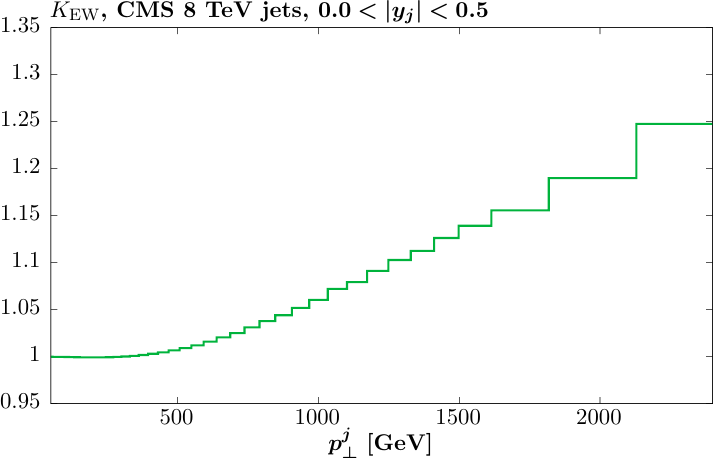}
\includegraphics[scale=0.6]{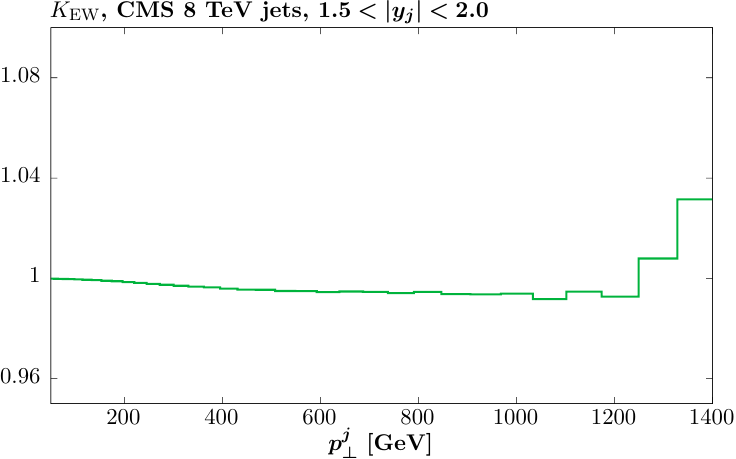}
\caption{\sf The calculated EW K-factors, corresponding to a selection of the CMS 8 TeV inclusive jet~\cite{CMS:2016lna} and dijet~\cite{CMS:2017jfq} data.}
\label{fig:kew}
\end{center}
\end{figure}

The K-factors are, as in~\cite{Harland-Lang:2017ytb,Bailey:2020ooq}, fitted to a 4--parameter polynomial in the logarithm of the binned variable. As the NNLO to NLO K-factors are expected to be smoothly varying functions, we argue this provides more control over any assessment of the impact of including NNLO theory in comparison to simply including the quoted MC errors in a bin--by--bin uncorrelated way.  Moreover, our default treatment of K-factors at  aN${}^3$LO is constructed by using using the NNLO (and NLO) K-factors in a manner that is based on these being smoothly varying functions, though is not reliant upon this. In Fig.~\ref{fig:kfs} we show the result of this, and the corresponding uncertainty, for a representative selection of the CMS 8 TeV inclusive jet~\cite{CMS:2016lna} and dijet~\cite{CMS:2017jfq} data (with both the LC and FC cases shown), and the fit is seen to work rather well. A moderate but clear difference in trend is seen between the LC and FC cases for the dijet distributions, as has been observed already in~\cite{Chen:2022tpk}. For the inclusive jet data, we take as our default renormalisation/factorisation scale $\mu = p_\perp^j$, however we will also consider $\mu = \hat{H_T}$, defined as the scalar sum of the transverse momentum of all partons in the event, see~\cite{Currie:2018xkj}. In Fig.~\ref{fig:kn3lo} we show the  NLO, NNLO and aN${}^3$LO QCD K-factors  corresponding to a selection of the CMS 8 TeV dijet~\cite{CMS:2017jfq} data, with the aN${}^3$LO QCD K-factors corresponding to the default dijet fit described in Section~\ref{sec:fitquality}. For concreteness we take the LC NNLO predictions here. The K-factors are determined as described in Section 7.1 of~\cite{McGowan:2022nag}.
 We can see that the predicted behaviour at aN${}^3$LO displays good perturbative convergence, in line with the lower order. Results for the inclusive jet case were presented in Section 7.2. of~\cite{McGowan:2022nag}, and so are not repeated here, but are also found to display similarly good perturbative convergence.

EW corrections are included as K-factors, and computed as in~\cite{Dittmaier:2012kx}. These include $O(\alpha \alpha_s)$ and $O(\alpha^2)$ and $O(\alpha\alpha_s^2)$ weak radiative corrections, that is they account for the dominant Sudakov logarithmic effect that becomes more significant at larger jet $p_\perp$ and dijet $m_{jj}$. Photon--initiated production is not included in these, but as discussed in~\cite{Cridge:2021pxm} these are negligible for jet production. The EW K-factors corresponding to a representative selection of the CMS 8 TeV inclusive jet~\cite{CMS:2016lna} and dijet~\cite{CMS:2017jfq} data are shown in Fig.~\ref{fig:kew}. We can see that in one bin these are as large as $\sim 10-20\%$ at larger values of the kinematic variable, $p_\perp^j$ or $p_\perp^{\rm av}$ in this case, but that for other rapidity bins they are significantly smaller. The magnitude of the EW corrections is moderately larger for the inclusive case, although not significantly so.

\subsection{Fit quality}\label{sec:fitquality}

We now consider the fit quality, i.e. $\chi^2$ per number of points, for global PDF fits including the jet and dijet data. As described above, the baseline data set is in both cases the same as MSHT20, but with the jet fit including 7 and 8 TeV inclusive jet data from ATLAS and CMS~\cite{ATLAS:2014riz,CMS:2014nvq,ATLAS:2017kux,CMS:2016lna}, and the dijet fit including instead the 7 and 8 TeV dijet data from ATLAS and CMS~\cite{ATLAS:2013jmu,CMS:2012ftr,CMS:2017jfq}. The inclusive jet fit then excludes the corresponding dijet data, and vice versa. For the sake of comparison, we will also consider `no jets/dijets' fits, which exclude both of these data sets, although continue to include the lower energy inclusive jet data from the LHC and Tevatron, again as described in the previous section.
These provide relatively limited constraints on PDFs.

In Table~\ref{tab:jets_chi2} we show the fit quality at both NNLO and aN${}^3$LO for these cases, while in Table~\ref{tab:jets_chi2nlo} we show results at NLO. In all cases, unless otherwise stated, NLO EW corrections are included for the jet and dijet data, with the treatment of the other data sets being as in MSHT20~\cite{Bailey:2020ooq,Cridge:2021pxm}. We note that in the original MSHT20aN3LO  study~\cite{McGowan:2022nag} the CMS 7 TeV inclusive jet data were taken with $R=0.5$, rather than $R=0.7$ that we now take, and moreover NLO EW corrections were omitted in the CMS 7 or 8 TeV inclusive jet data. Therefore our inclusive jet case now corresponds to the QCD only case of the recent aN${}^3$LO + QED study \cite{Cridge:2023ryv}. In Table~\ref{tab:jets_chi2} the fit qualities shown in bold are for data sets that are included in the corresponding fit, while the remainder are the predicted fit qualities from the resulting PDF set. In the aN${}^3$LO case, for the jet/dijet predictions  we apply the K-factors that are extracted from the corresponding fit with these data included, given these are not well determined (or determined at all in the dijet case) from the fit with these data sets excluded.

\begin{table}[t]
\scriptsize
    \centering
    \begin{tabular}{|c|c|c|c|c|c|c|c|c|c|}
    \cline{3-10}
    \multicolumn{2}{c|}{} & \multicolumn{3}{c|}{NNLO}&  \multicolumn{3}{c|}{aN${}^3$LO}&  \multicolumn{2}{c|}{aN${}^3$LO ($K_{\rm nnlo}$)}\\
    \hline
         & \multirow{1}{*}{$N_{\mathrm{pts}}$}  & \multicolumn{1}{c|}{No jets/dijets}  & \multicolumn{1}{c|}{Jets} & \multicolumn{1}{c|}{Dijets} & \multicolumn{1}{c|}{No jets/dijets} & \multicolumn{1}{c|}{Jets} & \multicolumn{1}{c|}{Dijets}& \multicolumn{1}{c|}{Jets}& \multicolumn{1}{c|}{Dijets}\\
            \hline
        ATLAS $7\ \text{TeV}$ jets~\cite{ATLAS:2014riz} &   140  &1.60& \bf{ 1.54} & 1.64 &1.72&  \bf{1.46}&   1.54&\bf{1.56}&1.44 \\
        CMS $7\ \text{TeV}$ jets~\cite{CMS:2014nvq} &   158  &1.39&  \bf{1.29} &    1.54&1.51 &\bf{1.32}&  1.34&\bf{1.33}&1.10  \\
        ATLAS $8\ \text{TeV}$ jets~\cite{ATLAS:2017kux}  &   171&2.02  & \bf{1.96} &   1.92 &2.03 & \bf{1.90}& 1.94&\bf{1.93}&1.83 \\
        CMS $8\ \text{TeV}$ jets~\cite{CMS:2016lna} &   174  &1.80 & \bf{1.83 } & 1.85&1.86 &\bf{ 1.80} &   1.74 &\bf{1.90}&2.06\\
               \hline
       Total (jets) & 643 &1.71& \bf{1.67} & 1.75 &1.79 &\bf{1.63} &  1.65&\bf{1.69}& 1.63\\
       \hline
                ATLAS $7\ \text{TeV}$ dijets~\cite{ATLAS:2013jmu} &   90  &1.08 &1.09 &  \bf{1.05}&1.13&1.13 &\bf{1.12}&1.13&\bf{1.12}\\
        CMS $7\ \text{TeV}$ dijets~\cite{CMS:2012ftr} &   54 &1.51 & 1.64 & \bf{1.44}&1.47& 1.47&\bf{1.40}&1.48&\bf{1.42} \\
        CMS $8\ \text{TeV}$ dijets~\cite{CMS:2017jfq} &   122&1.22  &  1.47 & \bf{1.22}&1.06 &1.01&\bf{0.86}&0.90&\bf{0.98} \\
        \hline
       Total (dijets) & 266&1.23 & 1.38 & \bf{1.21}&1.19&1.14  &  \bf{1.06}&1.10& \bf{1.12} \\
       \hline
        CMS $2.76\ \text{TeV}$ jets~\cite{CMS:2015jdl} &   81&\bf{1.28}  & \bf{1.25} &  \bf{1.32}&\bf{1.34} & \bf{1.37} &  \bf{1.32}&\bf{1.33}&\bf{1.42} \\
        ATLAS 8 TeV $Z$ $p_T$~\cite{ATLAS:2015iiu} &   104& \bf{1.75} & \bf{1.87} &  \bf{1.66}&\bf{0.99} & \bf{1.04} &  \bf{1.05} &\bf{1.37}&\bf{1.24}\\
        Differential $t\overline{t}$~\cite{CMS:2015rld,ATLAS:2015lsn,ATLAS:2016pal,CMS:2017iqf} &   54  &\bf{1.23}& \bf{1.10} & \bf{1.26}&\bf{1.11} & \bf{1.06} &  \bf{1.09} &
        \bf{1.06}&\bf{1.17}\\
             \hline
       Total &  - &\bf{1.15}& \bf{1.22} & \bf{1.15}&\bf{1.09} &  \bf{1.17} & \bf{1.09} &\bf{1.19}& \bf{1.11} \\
       \hline
    \end{tabular}
    \caption{\label{tab:jets_chi2}  $\chi^{2}$ breakdown per point for global fits at NNLO and aN$^{3}$LO including jet and dijet data, as described in the text. Data sets that are fit are shown in bold, while $\chi^2$ comparisons for data sets not included in the fit are not. The total number of points is 3891, 4543 and 4157 in the no jet/dijet, jet and dijet fits, respectively.}
\end{table}

\begin{table}
\scriptsize
    \centering
    \begin{tabular}{|c|c|c|c|}
    \cline{3-4}
    \multicolumn{2}{c|}{} & \multicolumn{2}{c|}{NLO}\\
    \hline
         & \multirow{1}{*}{$N_{\mathrm{pts}}$}   & \multicolumn{1}{c|}{Jets} & \multicolumn{1}{c|}{Dijets} \\
            \hline
        ATLAS $7\ \text{TeV}$ jets~\cite{ATLAS:2014riz} &   140  &\bf{1.60}& 1.83 \\
        CMS $7\ \text{TeV}$ jets~\cite{CMS:2014nvq} &   158  &\bf{1.37}&  1.81   \\
        ATLAS $8\ \text{TeV}$ jets~\cite{ATLAS:2017kux}  &   171&\bf{2.25}  & 2.34  \\
        CMS $8\ \text{TeV}$ jets~\cite{CMS:2016lna} &   174  &\bf{1.66} & 1.92   \\
               \hline
       Total (jets) & 643 &\bf{1.73}& 1.98  \\
       \hline
                ATLAS $7\ \text{TeV}$ dijets~\cite{ATLAS:2013jmu} &   90  &1.51 &\bf{1.12} \\
        CMS $7\ \text{TeV}$ dijets~\cite{CMS:2012ftr} &   54 &2.24 & \bf{1.70} \\
        CMS $8\ \text{TeV}$ dijets~\cite{CMS:2017jfq} &   122&7.84  &  \bf{5.27}  \\
        \hline
       Total (dijets) & 266&4.56 & \bf{3.14} \\
       \hline
       Total &  - &\bf{1.35}& \bf{1.42}  \\
       \hline
    \end{tabular}
    \caption{\label{tab:jets_chi2nlo}  $\chi^{2}$ breakdown per point for global fits at NLO including jet and dijet data, as described in the text. Data sets that are fit are shown in bold, while $\chi^2$ comparisons for data sets not included in the fit are not. The total number of points is 4534 and 4157 in the jet and dijet fits, respectively.}
\end{table}

Considering first the jet fits, we can see that the fit quality to the jet data improves from 1.73 to 1.67, and then to 1.63 from NLO to NNLO and aN${}^3$LO, respectively. That is, the fit quality improves with each order, as we would hope for. This improvement from NNLO to aN${}^3$LO is not observed in~\cite{McGowan:2022nag}, with the inclusion of NLO EW corrections for the CMS 7 and 8~TeV inclusive jet data sets and addition of the ATLAS 8~TeV jet data altering the behaviour. In any case, the improvement with each order is relatively mild, and even at aN${}^3$LO the fit quality remains poor. We recall that for the CMS 8 TeV jet data~\cite{CMS:2016lna} we include statistical correlations between bins; excluding these leads to a rather significant improvement in the fit quality for this data set, by e.g. $\sim 0.4$ per point at NNLO. For the ATLAS jet data, we find that the impact of the default correlations of the systematic errors is not negligible but relatively mild: at NNLO, the fit quality to the 7 (8) TeV data deteriorates by 0.2 (0.1) per point if these are used. The former result was already observed in~\cite{Bailey:2020ooq}. Also shown in the table is the result of a the aN${}^3$LO fit but with NNLO K--factors applied for the five classes of data sets described in~\cite{McGowan:2022nag} (namely all hadronic processes and dimuon production in semi--exclusive DIS). We can see that in this case the description of the jet data sets in fact deteriorates mildly with respect to the NNLO cases.  It is therefore not the case that a more precise aN${}^3$LO treatment of DIS data and DGLAP evolution alleviates some degree of tension with the inclusive jet data, as was also observed in \cite{McGowan:2022nag}. The inclusion of the aN${}^3$LO K-factors does provide sufficient freedom to give a net improvement in fit quality to jet data, but only a mild one.

 In the dijet fits we again see an improvement in the fit quality with increasing perturbative order, from 3.14 to 1.21 and 1.06 at  NLO, NNLO and aN${}^3$LO, respectively. However here the improvement is clearly more significant, with both the NNLO and aN${}^3$LO fits being rather good. Again we observe the aN${}^3$LO dijets fit with the NNLO K-factors has a very slightly worse fit quality relative to the  full aN${}^3$LO fit, though in the dijet case, still improved relative to the NNLO fit, indicating a preference for the known aN${}^3$LO information included in the fit.  For the CMS 8 TeV dijet data, we find that reverting to the  correlated systematic errors provided on the Hepdata repository, i.e. with apparently incorrect signs, gives a relatively small impact on the fit quality, with a deterioration of 0.02 per point at NNLO. We note that the combination of the relatively significant size of the 7 and 8 TeV inclusive jet data set with respect to the global data set (about $15\%$ in terms of the number of data points) and the significantly worse fit in this case results in the global fit quality at NNLO (aN${}^3$LO) being significantly improved in the dijet fit in comparisons to the jet fit, by 0.07 (0.08) per point.

At NNLO, if the K-factor points along with their quoted uncorrelated bin--by--bin statistical errors are used instead of the smooth fitting procedure described in the previous section, we find that the inclusive jet fit quality improves by $\sim 0.2$ per point. Such an improvement is not entirely surprising: the inclusive jet data are to a large extent dominated by systematic experimental errors (and their correlations) rather than statistical errors, and hence the fit quality is driven by the extent to which the theory can match these precise data. When the fit quality is relatively poor, including an additional source of uncorrelated uncertainty can therefore wash this out somewhat, even if the underlying K-factors are scattered by these errors, and lead to a better fit quality; effectively, the statistical errors on the data are being increased. Indeed, at NNLO and aN${}^3$LO  in the dijet case, where the fit quality is good, the difference between using the direct K-factor points and the fit is in general marginal. The one exception to this is when FC corrections are used for the CMS 8 TeV dijet data (discussed further below), for which the fit quality in fact deteriorates by $0.1$ ($0.05$) per point at NNLO (aN${}^3$LO) relative to the K-factor fit we use by default. Looking at Fig.~\ref{fig:kfs} there is perhaps some evidence that the scatter in the K-factors is somewhat larger than the quoted MC uncertainties, which may be driving this effect. 

Given these findings, we would therefore argue that  fitting the K-factors as smooth functions and including associated uncertainties  gives a clearer picture of the overall impact of such higher--order QCD corrections, when these are made available in this way. However, we find that the principle results, namely that the dijet fit quality is better, as well as the impact of EW corrections and the scale choice discussed further below, remain consistent irrespective of the treatment of the K--factors.

\begin{figure}
\begin{center}
\includegraphics[scale=0.6]{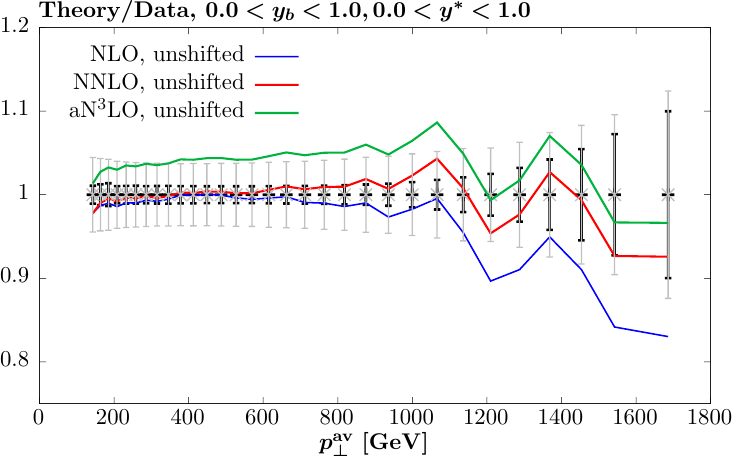}
\includegraphics[scale=0.6]{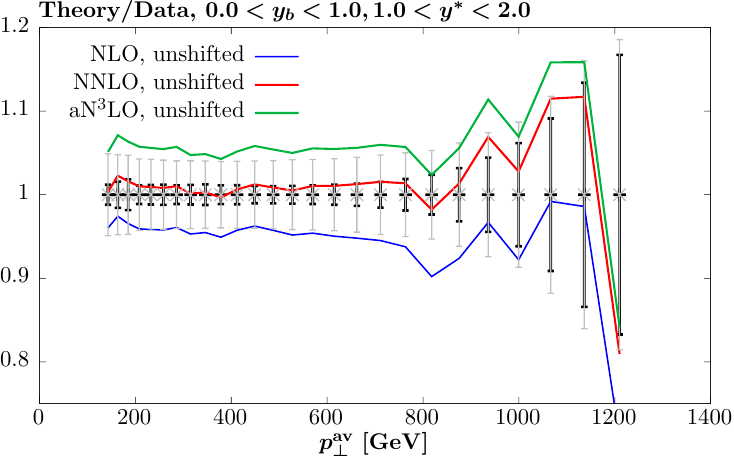}
\includegraphics[scale=0.6]{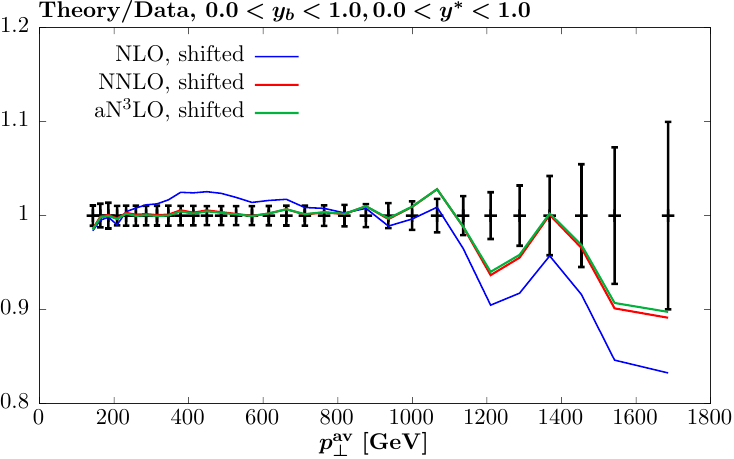}
\includegraphics[scale=0.6]{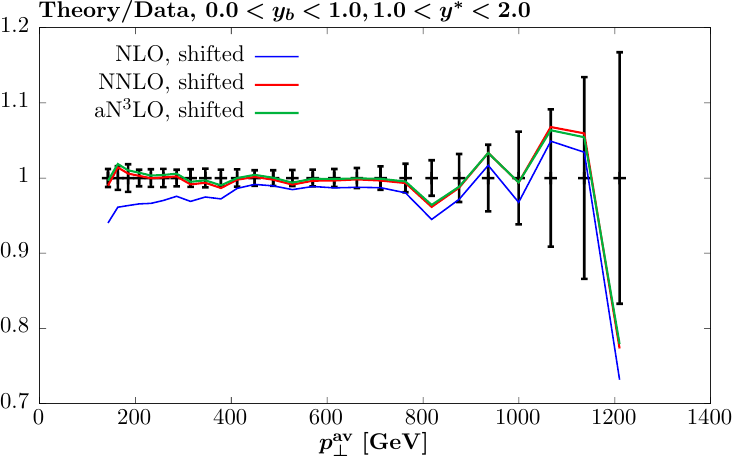}
\caption{\sf Theory/Data comparison for a selection of the CMS 8 TeV dijet~\cite{CMS:2017jfq} data, at different perturbative orders. The upper (lower) plots correspond to the result before (after) shifting by the correlated systematic errors. The black error bands are the purely statistical errors, while, the grey bands in the upper plot correspond to the statistical and total systematic error added in quadrature, shown for illustration.} 
\label{fig:cms8dat}
\end{center}
\end{figure}

The improvement from NLO to NNLO in the dijet fit is particularly dramatic, and we can see is driven by the description of the most constraining CMS 8 TeV dijet data, which at NLO is 5.27 per point, i.e. very poor indeed. At NNLO (aN${}^3$LO) on the other hand, this improves to 1.22 (0.86). Such a level of improvement from NLO to NNLO is not without precedent; in the MSHT20 analysis~\cite{Bailey:2020ooq}, for example, the description of the ATLAS high precision $W,Z$ data at 7 TeV~\cite{ATLAS:2016nqi} is 5.0 per point at NLO and 1.91 at NNLO. The latter admittedly remains a relatively poor fit, but the impossibility for purely NLO QCD to give anything other than an extremely poor description of such high precision data is clear. Indeed, in~\cite{McGowan:2022nag} the fit quality is found to improve again to 1.55 per point at aN${}^3$LO, with predicted K-factors that qualitatively  follow theoretical expectations~\cite{Chen:2021vtu}. We therefore view this improvement in the dijet case as again indicative of the failure of a NLO QCD analysis to match the increasing high precision and multi--differential data from the LHC, in this case the triple--differential CMS 8 TeV dijet data. 

To understand this further, in Fig.~\ref{fig:cms8dat} we show the theory/data for a selection of  the CMS 8 TeV dijet data, at the different perturbative orders. In the upper plots we show results before shifting by the correlated systematic errors, and while there is perhaps some hint that the NLO description will not be good (e.g. in the higher $p_\perp$ region) this is not completely clear. On the other hand, once we shift by the correlated systematic errors, as shown in the lower plot, we can clearly see that NLO theory cannot describe the shape of the corresponding distributions. It is therefore only after accounting for the systematic errors, and their appropriate correlations, that the poor fit quality in the NLO case becomes clear, and this is not completely evident by simply looking at data/theory comparisons by eye. We note that the mismatch of the shape when comparing data to theory at NLO  in the left of Fig.~\ref{fig:cms8dat} is roughly opposite to, and hence corrected by, the NNLO correction to the K-factor in Fig.~\ref{fig:kn3lo}, with the latter increasing with $p_\perp$. 

Turning now to the description of data not in the individual fits, at NNLO we find that the description of the jet/dijet data not included in the fit is $\sim 50$ points worse in both cases in comparison to the case where these are fit; the prediction of the jet data from the dijet fit is 0.08 per point worse, and the prediction of the dijet data from the jet fit is 0.17 per point worse. 
This indicates some degree of tension between the pulls of the two sets, even if it is not necessarily dramatic. At aN${}^3$LO on the other hand, the difference is significantly reduced, with 
 the description of the jet/dijet data not included in the fit being $\sim 15$ points worse in both cases. Therefore, although the description of the jet data remains relatively poor at this order, the degree of tension between the jet and dijet data is reduced. If only NNLO K--factors are used the description of the jet data in the dijet fit is $\sim 40$ worse, which might indicate a larger degree of tension in this case. On the other hand, the description of the dijet data by the jet fit is $\sim 8$ worse, i.e. less than in the full aN${}^3$LO  jets fit, so the picture is somewhat mixed.
 
It it also instructive to consider the fit quality to other LHC data sets that are known to be particularly sensitive to the high $x$ gluon, namely the ATLAS 8 TeV $Z$ $p_T$~\cite{ATLAS:2015iiu} data and the range of differential top quark pair production data from ATLAS and CMS~\cite{CMS:2015rld,ATLAS:2015lsn,ATLAS:2016pal,CMS:2017iqf} that are included in the baseline MSHT20 fit. At NNLO, we can see that the description of the $Z$ $p_T$ data improves from  1.87 to 1.66 per point, i.e. $\sim 22$ points in total, when the dijet rather than jet data are fit. This therefore indicates that some degree of the tension that is known to exist between the inclusive jet and $Z$ $p_T$ data (e.g. observed in \cite{Bailey:2020ooq,Thorne:2022abv}) is reduced when the dijet data are instead considered. The fit quality for the differential top data on the other hand is somewhat worse in the dijet fit, by 0.16 per point, indicating that these data are in somewhat larger tension with the dijet data than the inclusive jets. At aN${}^3$LO, on the other hand, the fit quality for the  $Z$ $p_T$ improves dramatically, as observed in~\cite{McGowan:2022nag}; it is close to 1 per point irrespective of whether the jet or dijet data are fit. This will be examined in more detail in Section~\ref{sec:zpt}. Similarly, any difference in the differential top data is reduced. This is indicative of the fact that the aN${}^3$LO tends to reduce such tensions in comparison to the NNLO case. If NNLO K--factors are used, we can see that the description of the $Z$ $p_T$ data remains  significantly improved with respect to NNLO, in both the case of the jet fit (again as observed in~\cite{McGowan:2022nag}) and the dijet fit. However, there is further room for improvement here, and we can see that for the dijet fit the description of the $Z$ $p_T$ data is $0.13$ per point lower. The description of the differential top data is $0.11$  per point worse, on the other hand. Therefore, the degree of tension is reduced purely by including aN${}^3$LO theory (absent the K--factors), but remains present between the $Z$ $p_T$ data and the inclusive jets, and the differential top data and the dijets.

\begin{table}
\scriptsize
    \centering
    \begin{tabular}{|c|c|c|c|c|c|c|c|}
    \cline{3-8}
     \multicolumn{2}{c|}{} & \multicolumn{2}{c|}{NLO}& \multicolumn{2}{c|}{NNLO}& \multicolumn{2}{c|}{aN$^{3}$LO}\\
     \hline 
         &$N_{\mathrm{pts}}$&$\mu = p_\perp^j$ & $\mu = H_T$ &$\mu = p_\perp^j$ & $\mu = H_T$ &$\mu = p_\perp^j$ & $\mu = H_T$ \\
            \hline
        ATLAS $7\ \text{TeV}$ jets~\cite{ATLAS:2014riz} &   140 &1.62&1.42&1.55&1.35&1.45 &1.31\\
        CMS $7\ \text{TeV}$ jets~\cite{CMS:2014nvq} &   118 &1.00&1.07&1.02 &1.24 & 1.00 &1.09\\
        ATLAS $8\ \text{TeV}$ jets~\cite{ATLAS:2017kux}  &   171 &2.21&1.84 &1.94 &1.87&1.88  &1.73\\
        CMS $8\ \text{TeV}$ jets~\cite{CMS:2016lna} &   174  &1.65&1.88& 1.83  &1.95 &1.80&2.06\\
               \hline
       Total (jets) & 603&1.68&1.60&1.64 &1.65& 1.58& 1.60 \\
             \hline
       Total &  4494 &1.33&1.32& 1.20 &1.20&1.15  &1.15 \\
       \hline
    \end{tabular}
    \caption{\label{tab:jets_chi2_scale}$\chi^{2}$ breakdown per point for global fits at NLO, NNLO and aN$^{3}$LO including jet data, with two choices of jet scale. For the CMS 7 TeV data,  NNLO K-factors for $\mu = H_T$ are only available for the more limited jet $p_\perp$ region as in the earlier analysis~\cite{CMS:2012ftr}, and hence a $p_\perp^j > 114$ GeV is imposed here, and for a direct comparison in the $\mu = p_\perp^j$ case in this table (but not elsewhere). }
\end{table}

We next consider in Table~\ref{tab:jets_chi2_scale} the impact of the scale choice in the jet fits at different perturbative orders. Namely, we consider changing from our default renormalisation/factorisation scale $\mu = p_\perp^j$, to  $\mu = \hat{H_T}$, defined as the scalar sum of the transverse momentum of all partons in the event, see~\cite{Currie:2018xkj}. For the CMS 7 TeV data, NNLO K-factors are only available for the $p_\perp$ binning of the earlier analysis~\cite{CMS:2012ftr}, i.e. not for the lower $p_\perp$ bins, rather than the updated analysis~\cite{CMS:2016lna}. In particular, these are only available for $p_\perp^j > 114$ GeV, rather than the lower value of  $p_\perp^j > 56$ GeV corresponding to the later analysis. We therefore impose this higher $p_\perp^j$ cut for the $\mu = \hat{H_T}$ case and, to maintain a direct comparison, the $\mu = p_\perp^j$ case presented in Table~\ref{tab:jets_chi2_scale}, although not elsewhere. It is argued in~\cite{Currie:2018xkj} that $\mu = \hat{H_T}$ is a more perturbatively stable scale choice than $\mu = p_\perp^j$, and indeed we can see that at NLO the fit quality is somewhat better for this choice, by 0.08 per point, i.e. $\sim 50$ points in $\chi^2$. However, at NNLO and aN${}^3$LO the difference is marginal, with the  $\mu = \hat{H_T}$ scale giving a slightly worse fit quality. In terms of the trend with increasing perturbative order we can see that while this improves order--by--order for $\mu = p_\perp^j$ (albeit to a relatively poor fit quality even at aN${}^3$LO), for $\mu = \hat{H_T}$ no clear trend of this sort is observed. Therefore, at NNLO and beyond there is no clear preference in terms of the fit quality between the two scale choices.

\begin{table}
\scriptsize
    \centering
    \begin{tabular}{|c|c|c|c|c|c|}
    \cline{3-6}
     \multicolumn{2}{c|}{} & \multicolumn{2}{c|}{NNLO}& \multicolumn{2}{c|}{aN$^{3}$LO}\\
     \hline 
         &$N_{\mathrm{pts}}$&Default & No EW. &Default & No EW \\
            \hline
        ATLAS $7\ \text{TeV}$ jets~\cite{ATLAS:2014riz} &   140 & 1.54& 1.48&1.46 &1.45\\
        CMS $7\ \text{TeV}$ jets~\cite{CMS:2014nvq} &   158 &1.29&1.24 & 1.32&1.31\\
        ATLAS $8\ \text{TeV}$ jets~\cite{ATLAS:2017kux}  &   171  &1.96&2.01&1.90&1.92\\
        CMS $8\ \text{TeV}$ jets~\cite{CMS:2016lna} &   174  & 1.83 &1.52 &1.80&1.60\\
               \hline
       Total (jets) & 643&1.67&1.57&1.63& 1.59 \\
             \hline
       Total &  4534 &1.22 &1.20&1.17 &1.17\\
       \hline
    \end{tabular}
    \caption{\label{tab:jets_chi2_EW}$\chi^{2}$ breakdown per point for global fits at NNLO and aN$^{3}$LO including jet data, with and without EW corrections included.}
\end{table}

In Tables~\ref{tab:jets_chi2_EW} and~\ref{tab:dijets_chi2_EW} we show the impact of excluding NLO EW corrections at both NNLO and aN$^{3}$LO in QCD, and for both the jet and dijet fits, respectively. In the dijet case, we can see that the fit quality with the EW corrections is improved by $\sim 0.22-0.23$ per point relative to the case without, i.e. by $\sim 60$ points in $\chi^2$, at both NNLO and aN${}^3$LO orders. The inclusion of EW corrections is therefore relatively significant in achieving an overall very good fit quality in both cases. However, in the jet case we can see that the fit quality at both orders in QCD actually deteriorates upon the inclusion of EW corrections, by 0.1 per point at NNLO and somewhat less (0.04 per point) at aN$^{3}$LO. The reason for this deterioration is unclear, but given these EW corrections should certainly be included we would in general expect them to improve the fit quality. Given this, this observation is arguably a further indication of an underlying issue in the fit to the inclusive jet data.  We note that if NLO EW corrections are excluded, then the fit quality to the jet data indeed gets mildly worse at aN$^{3}$LO in comparison to NNLO, consistent with the differing trend observed in~\cite{McGowan:2022nag} and discussed above.

\begin{table}
\scriptsize
    \centering
    \begin{tabular}{|c|c|c|c|c|c|}
    \cline{3-6}
     \multicolumn{2}{c|}{} & \multicolumn{2}{c|}{NNLO}& \multicolumn{2}{c|}{aN$^{3}$LO}\\
     \hline 
         &$N_{\mathrm{pts}}$&Default & No EW. &Default & No EW \\
            \hline
                  ATLAS $7\ \text{TeV}$ dijets~\cite{ATLAS:2013jmu} &   90  & 1.05&  1.33 &1.12&1.44\\
        CMS $7\ \text{TeV}$ dijets~\cite{CMS:2012ftr} &   54  & 1.44 &  1.59&1.40&1.56 \\
        CMS $8\ \text{TeV}$ dijets~\cite{CMS:2017jfq} &   122  &  1.22 & 1.44&0.86&1.06 \\
               \hline
       Total (dijets) & 266&1.21& 1.43&1.06& 1.29\\
             \hline
       Total &  4157  &1.15 &1.16&1.09 &1.11\\
       \hline
    \end{tabular}
    \caption{\label{tab:dijets_chi2_EW}$\chi^{2}$ breakdown per point for global fits at NNLO and aN$^{3}$LO including dijet data, with and without EW corrections included.}
\end{table}

Next, in Table~\ref{tab:dijets_chi2_FC} we examine the impact of including FC corrections at NNLO in case of the CMS 8 TeV dijet data~\cite{CMS:2017jfq}. We can see that this in fact leads to some moderate deterioration in the fit quality at NNLO, by 0.07 per point. This is not in general what we would expect, given these FC corrections should be included in the theoretical prediction. However, the deterioration corresponds to less than $1\sigma \sim 0.13  $ per point for this data set, so is of relatively limited statistical significance. Moreover, at aN$^{3}$LO  the difference is very mild, with the fit quality in the FC case in fact being slightly better. Interestingly, if NNLO K-factors are used, the difference is also rather less than at NNLO, although in this case the FC fit quality is slightly worse than the LC. This therefore indicates that difference between the NNLO and aN$^{3}$LO fits is not entirely driven by the freedom one has in the approximate  N$^{3}$LO K--factors to absorb in a large part differences between FC and LC results and NNLO; while the aN$^{3}$LO K-factor will be constructed from a different NNLO K-factor in the FC case, there remains a freedom in the corresponding nuisance parameters at this order, see~\cite{McGowan:2022nag}. However this does play some role, with the aN$^{3}$LO K-factor giving an improvement of $0.04$ per point in comparison to the NNLO K-factor, i.e. aN$^{3}$LO \,($K_{\rm nnlo}$), case. In other words going from the FC fit quality being 0.02 per point better in the full aN$^{3}$LO fit to 0.02 worse when the K-factors are fixed. 

We find that taking the FC result does not change any of the relevant conclusions above: that is, the fit quality in the dijet case remains significantly better at NNLO and aN$^{3}$LO, with a moderate improvement seen between these orders, while the impact of EW corrections remains that of improving the fit quality. We also note that there is some change in the fit quality to other data sets in the fit when a refit is performed with FC corrections included instead of LC, although this is rather small; the total improvement in $\chi^2$ upon refitting at NNLO is $\sim 4$ points, with $\sim 3$ of this being in the CMS 8 TeV dijet data. However we recall that FC corrections are not available for and thus also not included for any of the other jet data sets,  and so it is difficult to draw firm conclusions here.

\begin{table}
\scriptsize
    \centering
    \begin{tabular}{|c|c|c|c|c|c|c|c|}
    \cline{3-8}
     \multicolumn{2}{c|}{} & \multicolumn{2}{c|}{NNLO}& \multicolumn{2}{c|}{aN$^{3}$LO}&
     \multicolumn{2}{c|}{aN$^{3}$LO ($K_{\rm nnlo})$}\\
     \hline 
         &$N_{\mathrm{pts}}$&LC & FC &LC & FC&LC & FC \\
            \hline
        CMS $8\ \text{TeV}$ dijets~\cite{CMS:2017jfq} &   122  &  1.22 & 1.29&0.86&0.84&0.98&1.00 \\
               \hline
    \end{tabular}
    \caption{\label{tab:dijets_chi2_FC}$\chi^{2}$ breakdown per point for CMS 8 TeV dijet data~\cite{CMS:2017jfq} at  NNLO, aN$^{3}$LO and aN$^{3}$LO  (with NNLO K-factors) with leading (LC) and full colour (FC) corrections included at NNLO.}
\end{table}

We finish this section by commenting on the results of the earlier study in the NNPDF fit~\cite{AbdulKhalek:2020jut}, where largely the same 7 and 8 TeV jet and dijet data sets were fit at up to NNLO order in QCD. In terms of the inclusive jet fits, a completely direct comparison is not in all cases possible, as the baseline fit there only includes one rapidity bin in the ATLAS 7 TeV jet data~\cite{ATLAS:2014riz}, while for the CMS 7 TeV data the older analysis of~\cite{CMS:2012ftr} is used rather than the update~\cite{CMS:2014nvq} that we take by default, and for the CMS 8 TeV data statistical correlations are not included. Nonetheless, in terms of the overall trends we can observe some consistency but also some differences. In particular, the fit quality to the ATLAS jet data is relatively poor, similar to here for the 7 TeV and in fact rather worse at 8 TeV (although in the later NNPDF4.0 analysis~\cite{NNPDF:2021njg} this is significantly lower), while the fit quality to the CMS data is better, but consistent with what we find when we cut out the lower $p_\perp$ bins of the 7 TeV data and do not account for statistical correlations in the 8 TeV data. Although a fit to both 7 and 8 TeV data sets at NLO is not performed, when fit individually it is found in the NNPDF analysis that the fit quality to the inclusive jet data deteriorates from NLO to NNLO. This is not observed in this study, but given the various differences in the manner in which the inclusive jet data are treated and lack of a like--for--like comparison fit to both 7 and 8 TeV data, there are many potential reasons for this. The fit quality upon the inclusion of EW corrections is on the other found broadly to deteriorate, as we find.

For the dijet data, the fit quality to the CMS 8 TeV dijets~\cite{CMS:2017jfq} is in fact rather worse in \cite{AbdulKhalek:2020jut} than what is found here, being 1.58 per point in the default fit, in comparison to 1.22. The reason for this is not straightforward to determine, but is most likely due to the underlying differences between the MSHT and NNPDF fits, and most importantly the other data sets that are included in the fit, and the treatment of them. The fit quality at NLO is also found to be very poor (3.69 per point in a fit to 8 TeV dijet data) and to improve significantly upon the inclusion of NNLO theory. On the other hand, the fit quality is found to deteriorate upon the inclusion of EW corrections, contrary to our result here. Therefore, our results indeed display some similarities with the study of~\cite{AbdulKhalek:2020jut}, but also some differences. This highlights that the interpretation of a given set of data cannot always be evaluated in isolation, but is rather tied up with the other data in the PDF analysis and the manner in which these are treated, as well as other methodological differences. As an example, the treatment of the ATLAS 8 TeV $Z$ $p_T$~\cite{ATLAS:2015iiu} data, which impact on the gluon PDF in a similar (large) $x$ region, is rather different between the NNPDF and MSHT analyses, with NNPDF including somewhat less data and treating the uncertainties slightly differently, see Section~\ref{sec:zpt} for a detailed discussion.

In conclusion, we observe that the  fit quality to the dijet data is greatly improved with respect to the inclusive jet case at NNLO, and that this remains true at aN$^{3}$LO, where the fit quality to the inclusive jets remains relatively poor. This conclusion is unaffected if we take $\mu = H_T$ rather than $\mu = p_\perp^j$ for the renormalisation/factorisation scale in the inclusive jet theory, while the impact of FC corrections in the triple differential data is mild but not negligible. Moreover, we find that the fit quality to the inclusive jet data becomes worse when we include NLO EW corrections, in contrast to the dijet case, where we see an improvement, as we would in general expect. The underlying issue that is present in achieving a good fit to the jet data when binned inclusively is therefore not present in the dijet measurements, including the triple differential and highly constraining CMS 8 TeV data. This indicates that it may be preferable to include dijet rather than inclusive jet data in the fit, when both options are available. However when and if the full correlations between the two data sets become available, this conclusion will need to be reassessed.

\subsection{Impact on PDFs}\label{sec:PDFs}

\begin{figure}
\begin{center}
\includegraphics[scale=0.6]{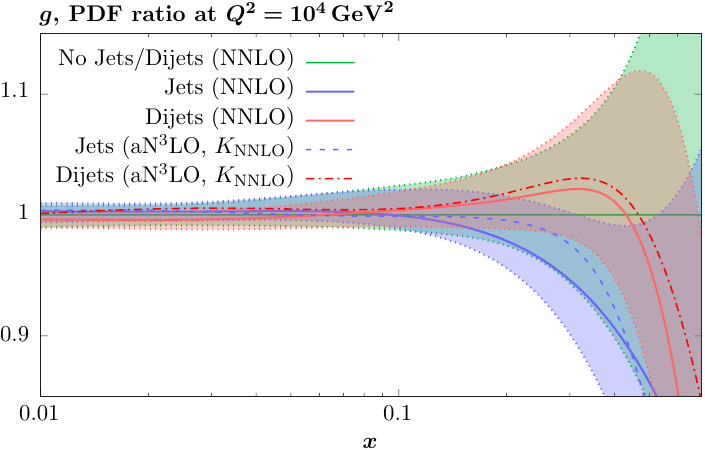}
\includegraphics[scale=0.6]{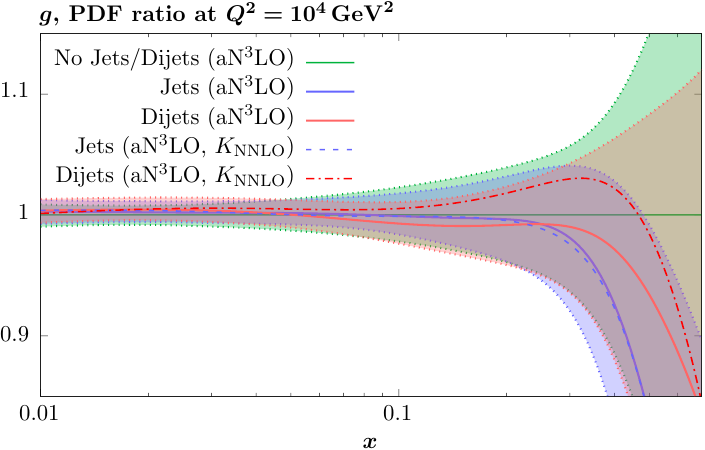}
\includegraphics[scale=0.6]{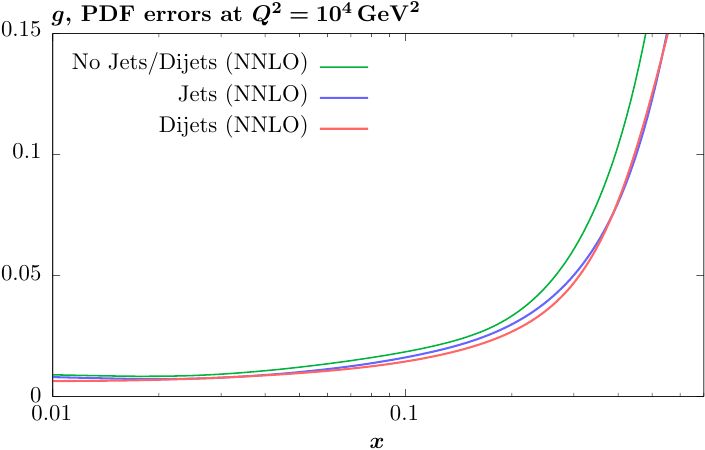}
\includegraphics[scale=0.6]{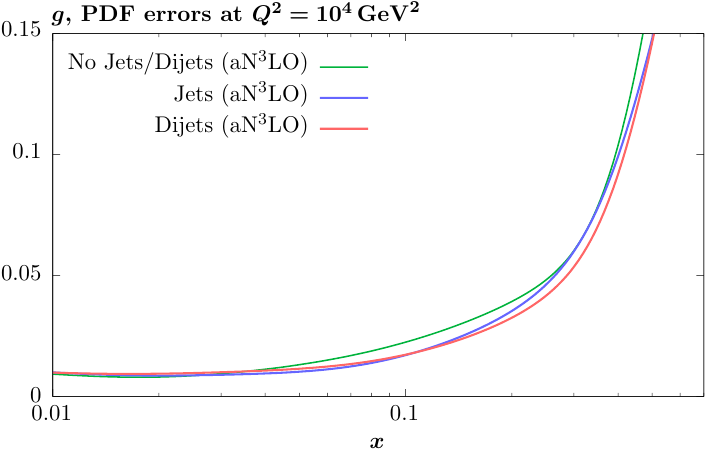}
\caption{\sf The gluon PDF resulting from the  jet and dijet fits, with respect to the no jets/dijets case. The left (right) plot corresponds to the NNLO (aN$^{3}$LO) case. The top plots show the PDF ratio, including the 68\% C.L. PDF errors, while the bottom plots show the symmetrised errors. Also shown in both top plots are jet and dijet cases at aN$^{3}$LO, but with NNLO K-factors.}
\label{fig:gluon}
\end{center}
\end{figure}

We next consider the impact of the fits described in the previous section on the PDFs. We begin in Fig.~\ref{fig:gluon} with the gluon PDF, as we expect jet and dijet data to have the largest impact in this case, in particular at high $x$. We can see from the top left plot that at NNLO, while they are consistent within errors, the jet and dijet data have somewhat different pulls on the gluon, with the dijet data preferring a somewhat larger gluon. This is consistent with the reasonable degree of tension observed in the two fits in Table~\ref{tab:jets_chi2}, as well as with the fact that the dijet fit give a rather better description of the ATLAS $Z$ $p_T$ data~\cite{ATLAS:2015iiu}, which is found to prefer a larger high $x$ gluon in the MSHT20 fit~\cite{Bailey:2020ooq,Thorne:2022abv,Jing:2023isu}. At aN$^{3}$LO, on the other hand, the pull on the gluon is more consistent, which is again as expected from the smaller degree of tension observed between the two fits in Table~\ref{tab:jets_chi2}. 

To investigate this further, we also show in Fig.~\ref{fig:gluon} (right) the impact on the gluon at  aN$^{3}$LO, but with NNLO K-factors. In this case the jet and dijet results show larger deviations at high $x$. Indeed, in the left plot we also include the same  $K_{\rm nnlo}$ curves for comparison, and the difference between the jet and dijet case with NNLO K-factors is more similar to that at NNLO. For the jet fit, the majority of the change in going to aN$^{3}$LO can be seen to come from the other aN$^{3}$LO information in the fit, as in the right figure we can see that the full aN$^{3}$LO and $K_{\rm nnlo}$ results are very similar. For the dijet fit, the trend is different, and the inclusion of aN$^{3}$LO K-factors does have some impact on the gluon. Overall, while we  can see in the left figure that the aN$^{3}$LO (with NNLO K-factor) curves are somewhat closer than the pure NNLO case,  the impact of including aN$^{3}$LO K-factors is more significant, bringing the jet and dijet gluons into closer agreement. Thus, to a large extent the reduction in the (mild) tension between the jet and dijet fits with respect to the high $x$ gluon at aN$^{3}$LO is due to the freedom allowed by the parameterised K--factors at this order, and results in a gluon in the dijet fit that lies closer to the jet case.

\begin{figure}[t]
\begin{center}
\includegraphics[scale=0.6]{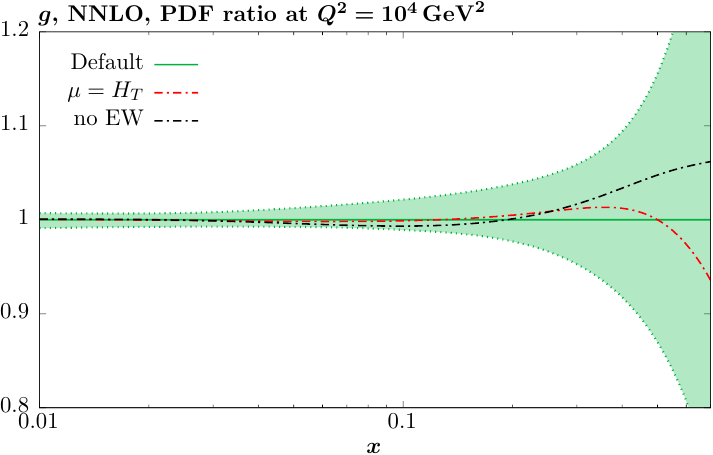}
\includegraphics[scale=0.6]{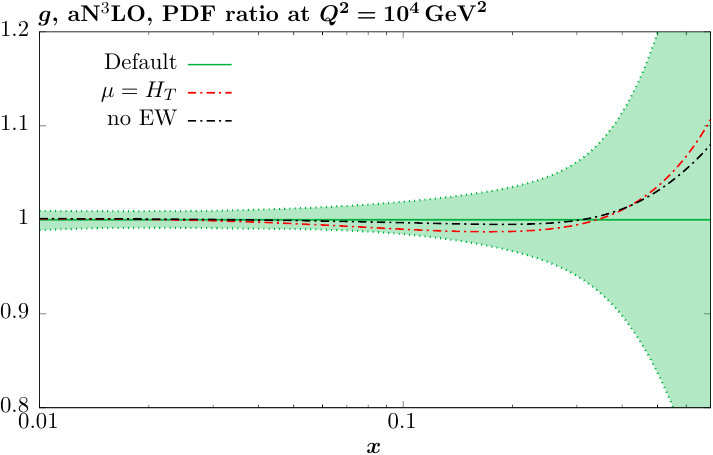}
\caption{\sf The impact of excluding EW corrections and taking $\mu = H_T$ on the gluon PDF resulting from the jet NNLO fits, with respect to the default ($\mu = p_\perp^j$, with EW corrections) fit. The 68\% C.L. PDF errors are shown for the baseline fit only.}
\label{fig:gluonewht}
\end{center}
\end{figure}

\begin{figure}[t]
\begin{center}
\includegraphics[scale=0.6]{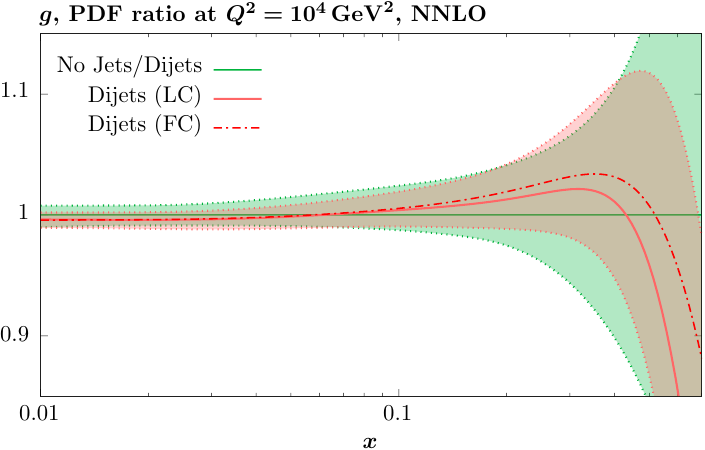}
\includegraphics[scale=0.6]{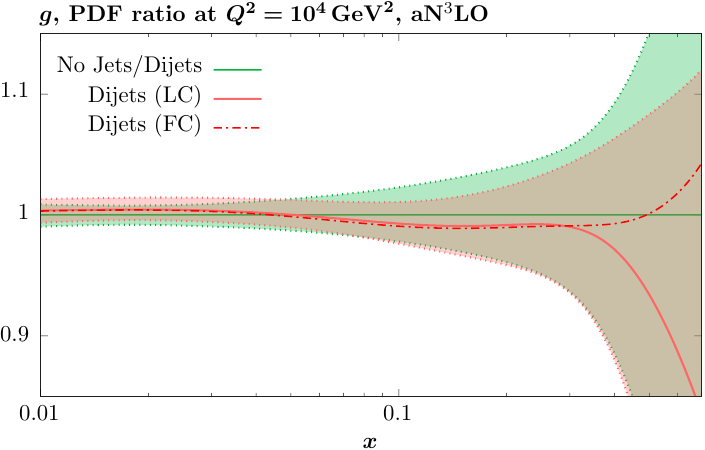}
\caption{\sf The impact of including full colour (FC) corrections at NNLO for the CMS 8 TeV dijet data~\cite{CMS:2017jfq} in the dijet fits at NNLO and aN$^{3}$LO in QCD, with respect to the default fit to both data sets. The 68\% C.L. PDF errors are shown for the baseline and LC fits only.}
\label{fig:gluonfc}
\end{center}
\end{figure}

In terms of the PDF uncertainties, shown in the lower plots after symmetrising, a clear but moderate reduction with respect to the no jets/dijets fit is observed. This reduction is comparable between the jet and dijet fits, but overall the dijet fits give a larger reduction, at both orders. While this relative improvement is quite small, it is worth noting that in terms of the bare number of data points, the dijet data are over a factor of 2 less. Indeed, most of the constraint comes from the CMS 8 TeV dijet data, which has a factor of $\sim 5$ less data points. Although such a measure only provides a rough guide, it is clearly notable that even given this the reduction of the gluon PDF uncertainty is  slightly greater for the dijet fit; for a larger data set we may expect further improvements.

In Fig.~\ref{fig:gluonewht} we show the impact of excluding NLO EW corrections and separately of using $\mu = H_T$ rather than $\mu = p_\perp^j$ for the renormalisation/factorisation scale in the inclusive jet case, corresponding to a subset of the fits in Tables~\ref{tab:jets_chi2_scale}, \ref{tab:jets_chi2_EW},\ref{tab:dijets_chi2_EW}. We can see that the effects are small and always within the PDF uncertainties, but not entirely negligible, such that the central values can approach close to the edge of the PDF uncertainty band of the baseline fit in some regions. There is no particular trend for a reduced impact at aN$^{3}$LO.

\begin{figure}[t]
\begin{center}
\includegraphics[scale=0.6]{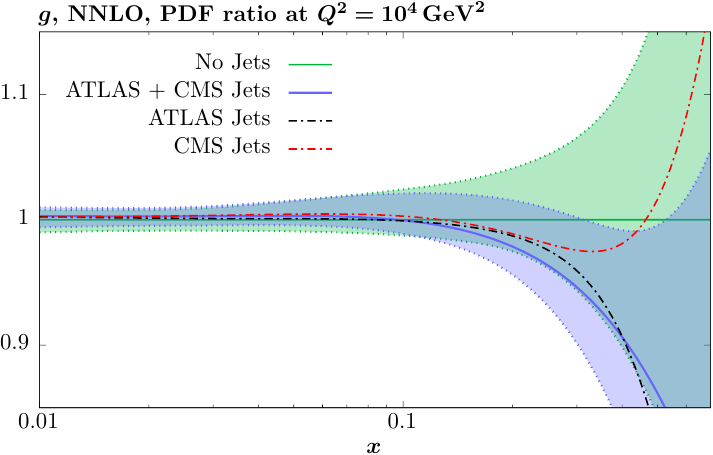}
\includegraphics[scale=0.6]{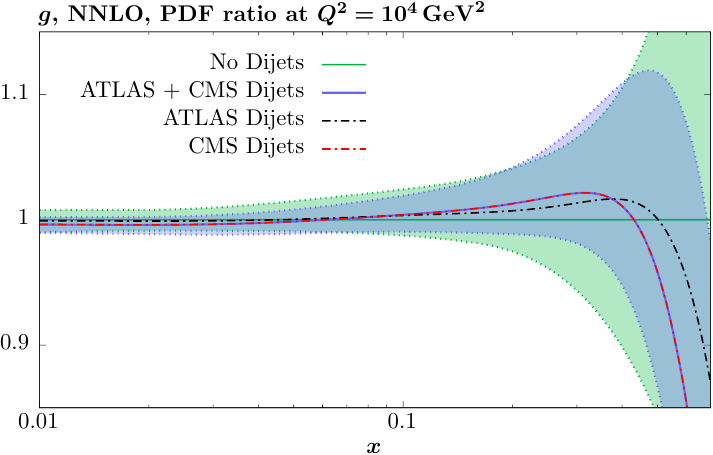}
\includegraphics[scale=0.6]{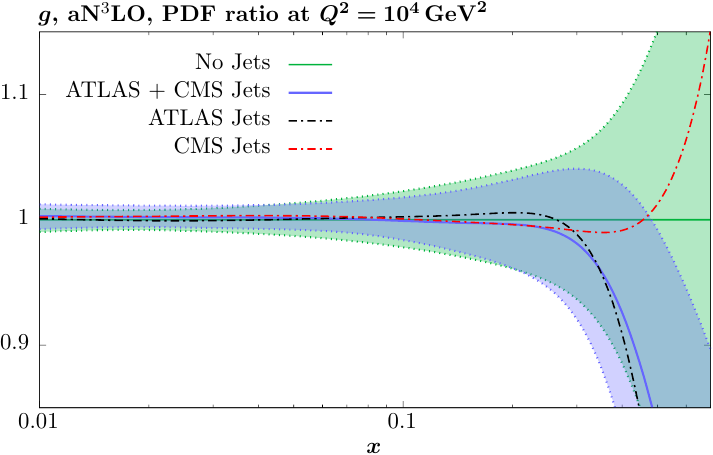}
\includegraphics[scale=0.6]{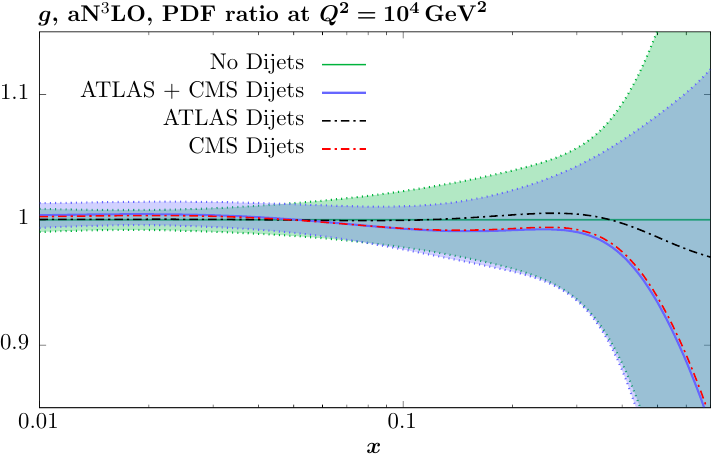}
\caption{\sf The impact of including only the ATLAS or CMS data on the gluon PDF in the jet and dijet fits at NNLO and aN$^{3}$LO in QCD, with respect to the default fit to both data sets. The 68\% C.L. PDF errors are shown for the baseline fit only.}
\label{fig:gluonatcms}
\end{center}
\end{figure}

In Fig.~\ref{fig:gluonfc} we show the impact of including FC corrections at NNLO to the CMS 8 TeV dijet data, in the dijet fits at NNLO and aN$^{3}$LO. The impact is mild and well within PDF uncertainties, though not entirely negligible at high enough $x$. 

\begin{figure}
\begin{center}
\includegraphics[scale=0.6]{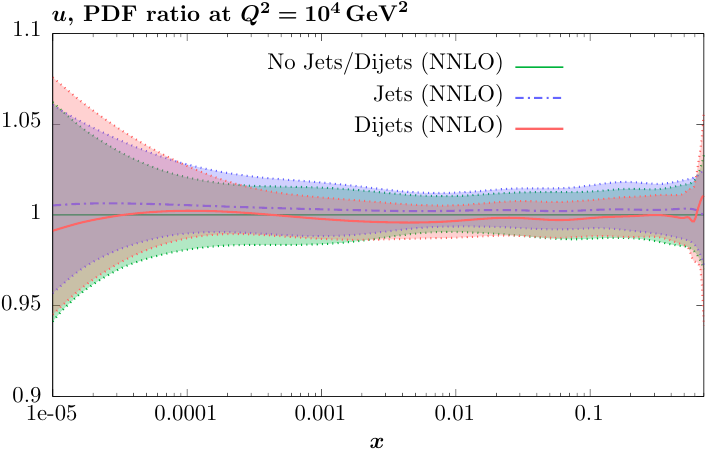}
\includegraphics[scale=0.6]{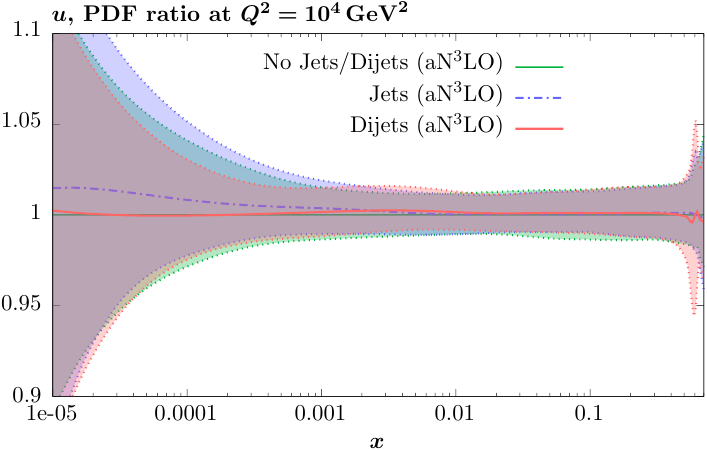}
\includegraphics[scale=0.6]{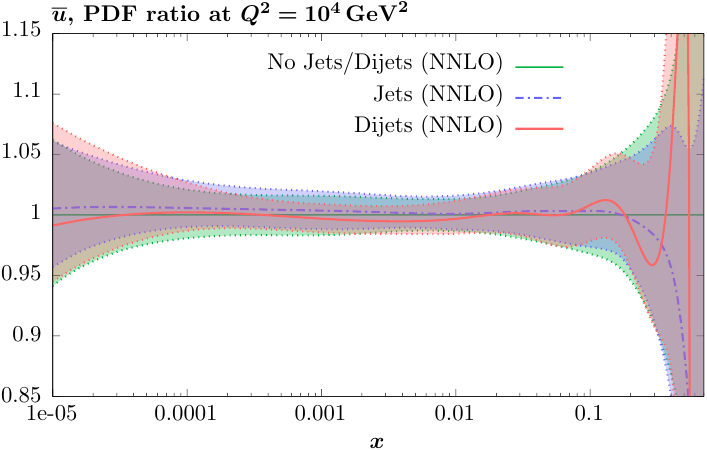}
\includegraphics[scale=0.6]{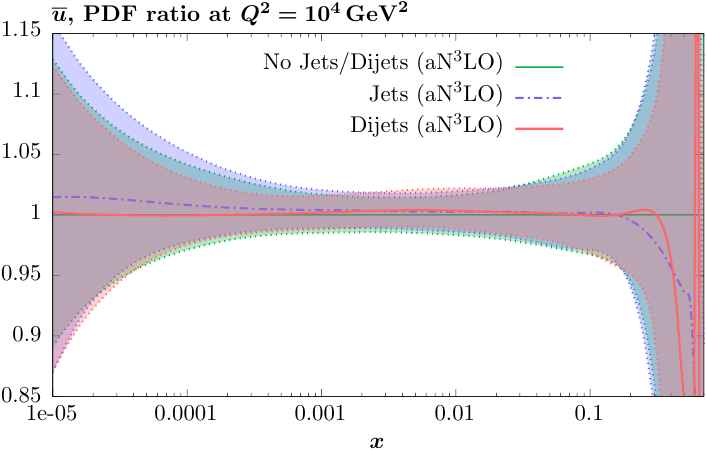}
\includegraphics[scale=0.6]{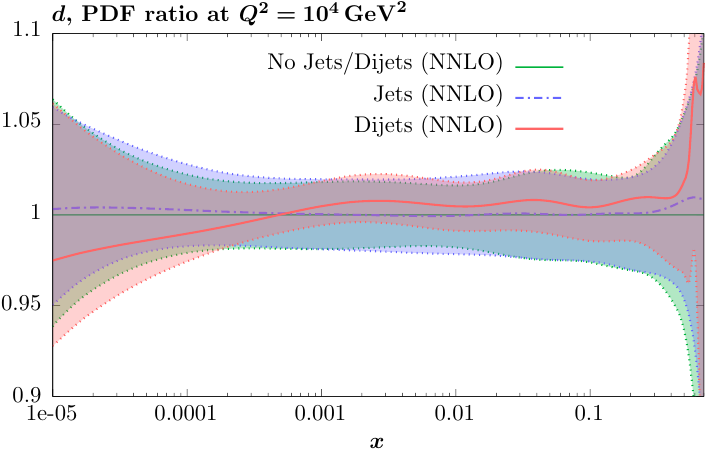}
\includegraphics[scale=0.6]{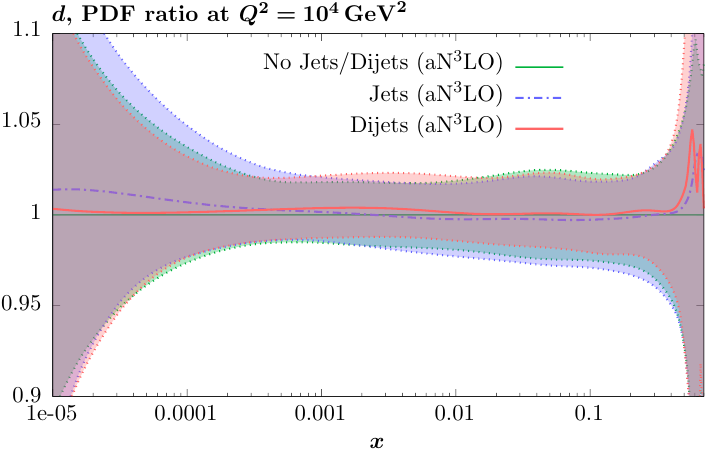}
\includegraphics[scale=0.6]{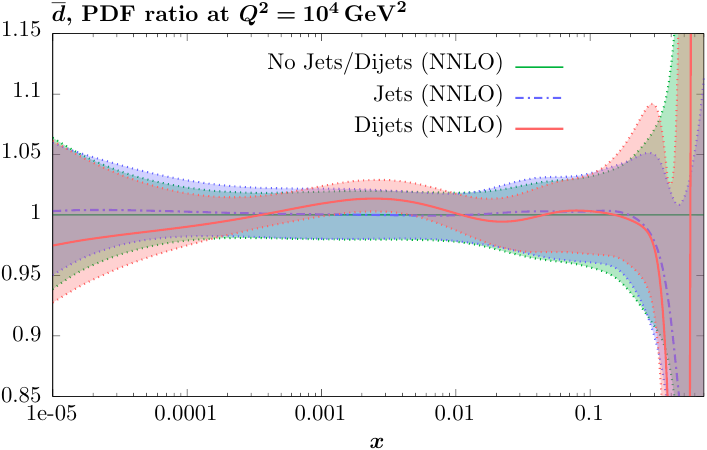}
\includegraphics[scale=0.6]{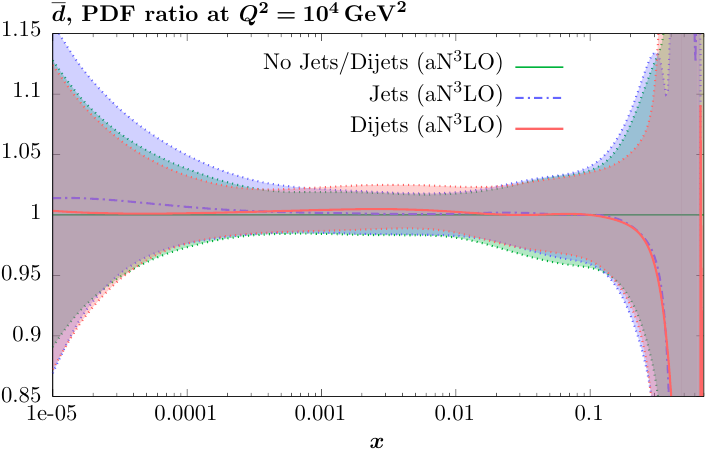}
\includegraphics[scale=0.6]{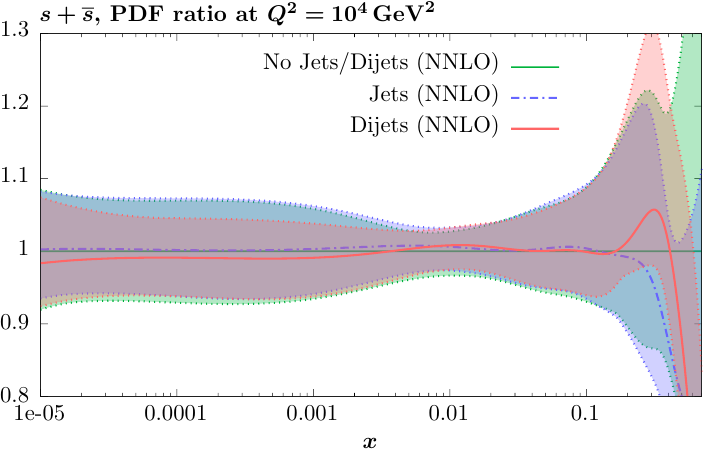}
\includegraphics[scale=0.6]{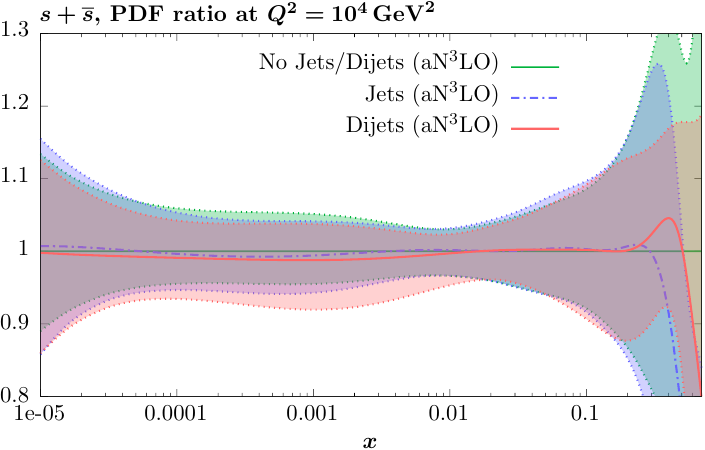}
\caption{\sf The quark PDFs resulting from the  jet and dijet fits, with respect to the no jet case, with 68\% C.L. PDF errors given. The left (right) plots show the NNLO (aN$^{3}$LO) fits.}
\label{fig:otherpdfs}
\end{center}
\end{figure}

In Fig.~\ref{fig:gluonatcms} we show the impact of only fitting the ATLAS or CMS jet/dijet data, to examine any difference in pull between these. In the left plots the results of the jet fits are shown, and we can see that the ATLAS and CMS data do indeed show some difference in pulls on the higher $x$ gluon (a similar effect was observed in~\cite{Harland-Lang:2017ytb} at NNLO), indicating a degree of tension. This remains true at  aN$^{3}$LO, as see in the bottom left plot.  In the right plots the corresponding dijet cases are shown, and here we can see that the ATLAS and CMS data tend to pull in a more similar  direction, indicating less of a degree of tension. Overall, the CMS curve follows the baseline (ATLAS + CMS) curve very closely, due to the fact that the 8 TeV triple differential CMS data are driving the fit in this case.

Finally, in Fig.~\ref{fig:otherpdfs} we show the impact on the quark flavour decomposition, at both NNLO (left figures) and aN$^{3}$LO (right figures). Overall, the impact on these PDFs is indeed less than in the gluon case at high $x$. However, interestingly at NNLO the dijet fit leads to some rearrangement of the quark flavour decomposition at intermediate $x$. The down quark and antiquark are particularly affected, and indeed have rather smaller uncertainties. The reason for this is not obvious, but appears to be simply due to the fact that the location of the global PDF minimum in the dijet case is such that the down type quarks are rather better constrained. At aN$^{3}$LO, on the other hand, this effect is rather smaller, although the uncertainty on the $d$  remains somewhat smaller in comparison to the jet, and no jet/dijet, cases. We note that the aN$^{3}$LO with NNLO K--factors gives rather similar results in comparison to the full aN$^{3}$LO fit, and hence are not shown.

%\newpage

\section{ATLAS 8 TeV $Z$ $p_T$ Data at NNLO and aN$^{3}$LO}\label{sec:zpt}

\subsection{Data and Theory}\label{sec:zptdatatheory}

A further source of constraints on the high $x$ gluon in the MSHT20 NNLO \cite{Bailey:2020ooq} and aN${}^3$LO \cite{McGowan:2022nag} PDFs is the ATLAS 8 TeV measurement of the dilepton transverse momentum distribution~\cite{ATLAS:2015iiu}, which we refer to for brevity as the $Z$ $p_T$ data set, although it extends beyond the peak region. Different groups have reported slightly different impacts of these data \cite{Hou:2019efy,NNPDF:2021njg}, though different amounts of data are included. In the MSHT20 NNLO and aN${}^3$LO PDF fits we choose to fit the maximum amount of data possible for this data set and include the absolute cross sections. Therefore we include the data double-differential in the lepton pair transverse momentum, $p_T^{ll}$, and rapidity, $y_{ll}$, in the $Z$ peak mass bin ([66,116]~GeV). For all other mass bins we take the available single differential distributions in $p_T^{ll}$. We include a $p_T^{ll}>30\,{\rm GeV}$ cut in order to exclude the region influenced by transverse momentum resummation and non-perturbative corrections. Finally, no maximum $p_T^{ll}$ cut is applied as electroweak corrections (which are relevant here) are included as provided in \cite{Boughezal:2017nla}. We therefore include a total of 104 data points, more than included by default by other global fitting groups. Our theory predictions are calculated using {\tt MCFM} \cite{Campbell:2002tg} with {\tt APPLgrid} at NLO \cite{Carli:2010rw} with NNLO K-factors determined from
{\tt NNLOJET} \cite{Gehrmann-DeRidder:2016cdi,Bizon:2018foh} and using the factorisation scale as the transverse mass of the vector boson, as is standard. Further details are given in \cite{Bailey:2020ooq} and in the next section. For the baseline data set we choose, for simplicity and ease of comparison with earlier studies, to fit to the same inclusive jet data described in the previous sections. However, we may expect broadly similar conclusions if instead the dijet baseline data set were fit.

\subsection{Fit Quality}\label{sec:zptfitquality}

It was noted in \cite{McGowan:2022nag} and earlier in Section~\ref{sec:fitquality}  (see Table~\ref{tab:jets_chi2}) that there was a significant improvement in the fit quality of these ATLAS 8~TeV $Z$ $p_T$ data at aN${}^3$LO relative to NNLO, with the latter quite poorly fit whilst the former was able to fit these data without issues. This suggested a clear preference in these precision data for aN${}^3$LO QCD corrections to the PDFs. The significant improvement at aN${}^3$LO relative to NNLO was also observed to be largely independent of the choice of inclusive jet or dijet data in the fit and therefore unconnected to potential tensions between these data sets. Nonetheless, it was noted in Section~\ref{sec:fitquality} that in absolute terms the fit quality of the $Z$ $p_T$ data set was notably better in the dijet fit at NNLO, indicating greater tension between the $Z$ $p_T$ and inclusive jet data than the dijet data, also observed in the pulls in Fig.~\ref{fig:gluon} (upper left). This difference is absent at aN${}^3$LO. 
In this section, we therefore investigate these ATLAS 8~TeV $Z$ $p_T$ data further in the context of the updated NNLO and aN${}^3$LO baseline PDFs used in this paper with the addition of the ATLAS 8~TeV inclusive jet data relative to our MSHT20 \cite{Bailey:2020ooq} baseline. 

Table~\ref{tab:Zptcut_chisqs} provides the fit qualities in terms of the $\chi^2$ per datapoint at NNLO and aN${}^3$LO for these data, first in our default setup with a $p_T^{ll}$ cut removing the resummation region below $30\,\rm{GeV}$. The fit qualities of 1.87 and 1.04 for the NNLO and aN${}^3$LO fits respectively follow the pattern previously noted of a poor fit quality at NNLO being significantly ameliorated at aN${}^3$LO. In order to investigate whether this improvement at aN${}^3$LO is associated with a particular part of the $p_T^{ll}$ spectrum, and thereby assess any potential limitations of the NNLO fit to these data, we systematically increase the $p_T^{ll}$ minimum cut incrementally and analyse the fit quality of these data at each order. Table~\ref{tab:Zptcut_chisqs} shows a gradual improvement of the fit quality of the NNLO fit from 1.87 (default $p_T^{ll} > 30\,{\rm GeV}$) to 1.24 (for $p_T^{ll} > 105\,{\rm GeV}$). By way of comparison there is also a small improvement in the fit quality of the aN${}^3$LO fit from 1.04 to 0.83 over the same range. Nonetheless the fit quality at NNLO remains poor and is always considerably worse than at aN${}^3$LO. There is therefore no clear evidence of issues associated with any particular part of the $p_T^{ll}$ spectrum at NNLO. In particular the fit remains comparatively poor even with quite high values of $p_T^{ll}$ cut. Thus, there is no evidence that sensitivity to resummation or other effects at low $p_T^{ll}$ higher than our default $p_T^{ll} > 30\,{\rm GeV}$ cut is responsible for the poor fit quality at NNLO. In addition, the final column of Table~\ref{tab:Zptcut_chisqs} shows the effect of imposing a maximum $p_T^{ll} < 150\,{\rm GeV}$ cut, which removes the larger $p_T^{ll}$ region where there is sensitivity to electroweak corrections (though these are included for these data). A poor fit quality of 1.91 remains at NNLO, and is again significantly improved to 1.08 at aN${}^3$LO, indicating sensitivity to electroweak or other effects in the high $p_T^{ll}$ region are also not responsible for the poor fit quality at NNLO. 

There is also evidence of reduced tension of the $Z$ $p_T$ data with other data sets in the global fit at aN${}^3$LO in comparison with those at NNLO. The rest of the global fit data at NNLO changes by $\Delta \chi^2 = -35.9$ upon the removal of the ATLAS 8~TeV $Z$ $p_T$ data\footnote{Note this is slightly different to that reported in \cite{Bailey:2020ooq} due to the addition of the ATLAS 8~TeV inclusive jets data to the baseline here, as well as other minor alterations.}. This reflects the tensions of these data at NNLO with a range of other data sets, as shown further in Table~16 of \cite{Bailey:2020ooq}. In contrast the corresponding change at aN${}^3$LO of the remainder of the data once the ATLAS 8~TeV $Z$ $p_T$ is removed is $\Delta\chi^2 = -22.0$, demonstrating a reduced sensitivity to the presence of these data at aN${}^3$LO and indicative of reduced tensions between it and other data in the aN${}^3$LO fit. Table~\ref{tab:Zptremoval_chisqs} demonstrates how this improvement is spread across a selection of data sets in the global fit. The data sets showing the greatest absolute improvements in $\chi^2$ once the ATLAS 8~TeV $Z$ $p_T$ data are removed in the NNLO fit are the NMC deuteron data, the HERA $e+p$ NC 920~GeV data, the ATLAS 7~TeV inclusive jets data, the ATLAS 7~TeV precision $W,Z$ data, the ATLAS 8~TeV $Z$ data and the ATLAS 8~TeV inclusive jets data. These all show reduced $\Delta\chi^2$ improvements, or in some cases no improvement, upon removal of the $Z$ $p_T$ data at aN${}^3$LO. Similar reduced tensions were also noted in \cite{Jing:2023isu}, for example see Fig.~17. Consistency is observed with the trends seen there with the NMC deuteron, HERA $e+p$ NC 920~GeV data and the ATLAS 7~TeV inclusive jets data being amongst those shown to oppose the ATLAS 8~TeV $Z$ $p_T$ pull on the gluon. We therefore observe the improvement in the fits, though often mild, of several of the inclusive jet data sets when the $Z$ $p_T$ data are removed, the same is true to a lesser extent for some of the $t\bar{t}$ data sets, indicating some tension between these data sets and the ATLAS 8~TeV $Z$ $p_T$ data.

\begin{table}
\fontsize{9.5}{12}\selectfont 
 \renewcommand\arraystretch{0.99} 
\centering \hspace{-0.1cm} \begin{tabular}{|>{\centering\arraybackslash}m{1.75cm}|>{\centering\arraybackslash}m{1.9cm}|>{\centering\arraybackslash}m{0.75cm}|>{\centering\arraybackslash}m{0.75cm}|>{\centering\arraybackslash}m{0.75cm}|>{\centering\arraybackslash}m{0.75cm}|>{\centering\arraybackslash}m{0.75cm}|>{\centering\arraybackslash}m{0.75cm}|>{\centering\arraybackslash}m{3.5cm}|}
\hline
& \multicolumn{7}{>{\centering\arraybackslash}m{9.2cm}|}{$p_T^{ll}$ minimum cut (GeV)} & $p_T^{ll}$ maximum cut (GeV) \\ \hline
Fit Order & Default (30) & 45 & 55 & 65 & 75 & 85 & 105 & 150 \\ \hline
NNLO & 1.87 & 1.73 & 1.72 & 1.47 & 1.45 & 1.47 & 1.24 & 1.91 \\ 
aN${}^3$LO & 1.04 & 0.97 & 1.03 & 0.86 & 0.88 & 0.71 & 0.83 & 1.08\\ \hline
$N_{\rm pts}$ & 104 & 88 & 77 & 66 & 55 & 44 & 33 & 82 \\ \hline 
\end{tabular}
\vspace{0.1cm}
\caption {Fit qualities, i.e. $\chi^2/N_{\rm pts}$, for NNLO and aN${}^3$LO MSHT PDF fits varying the $p_T^{ll}$ cut applied for the ATLAS 8~TeV $Z p_T$ data.}
\label{tab:Zptcut_chisqs}
\end{table}

In addition to comparing the different fit qualities with and without the ATLAS 8~TeV $Z$ $p_T$ data set, or with differing minimum and maximum $p_T^{ll}$ cuts applied, we can also compare the treatments of different global fitting groups of these data. In MSHT we by default include 104 datapoints for the ATLAS 8~TeV $Z$ $p_T$ data set. As outlined in the previous section, this corresponds to including all data with $p_T^{ll} > 30\, {\rm GeV}$ with single differential results in the dilepton transverse momentum $p_T^{ll}$ for the mass bins not including the $Z$ boson invariant mass peak - i.e. [12,20], [20,30], [30,46], [46,66] and [116,150] where $[a,b] = a\, {\rm GeV} <  m_{ll} < b\, {\rm GeV}$. In addition, double differential results in $p_T^{ll}$ and the rapidity of the dilepton pair $y_{ll}$ are utilised in the [66,116] mass bin incorporating the $Z$  mass peak. In contrast, NNPDF and CT take slightly different data subsets. NNPDF (from  3.1 onwards) additionally cut datapoints with $p_T^Z > 150 \,{\rm GeV}$ in the $Z$ mass bin [66,116], resulting in 12 fewer datapoints, the justification being to remove sensitivity to large $p_T^Z$ where electroweak corrections become large~\cite{NNPDF:2021njg,NNPDF:2017mvq}. CT18 apply the same large $p_T^{ll}$ cut as NNPDF, however they also cut the region $p_T^{ll} < 45 \,{\rm GeV}$ (compared to $p_T^{ll} < 30\, {\rm GeV}$ in MSHT20 and NNPDF4.0), to be more conservative in reducing sensitivity to the low $p_T^{ll}$ resummation region. On top of this they only fit the data in the mass bins [46,66], [66,116] and [116,150] (with the argument being to eliminate lower $M_{ll}$ bins where higher order corrections are potentially more significant) and treat the central $Z$ peak mass bin only single differentially in $p_T^{ll}$, i.e. not including the rapidity dependence. This therefore results in significantly fewer datapoints, with only 27 ultimately being fitted. There are additionally differences in the treatment of the uncertainties which also impact the fit quality and pulls on the PDFs observed, with uncorrelated uncertainties included in the CT and NNPDF cases, as discussed in \cite{Bailey:2020ooq}. In the latter NNPDF case the size of these is significantly larger than the quoted MC uncertainties on the K-factors and the improvement in the fit quality was found to be dramatic. These other differences will not be further examined in this work. 

\begin{table}[t]
\fontsize{10}{12}\selectfont 
 \renewcommand\arraystretch{0.99} 
\centering \hspace{-0.1cm} \begin{tabular}{|>{\centering\arraybackslash}m{7.0cm}|>{\centering\arraybackslash}m{1.0cm}|>{\centering\arraybackslash}m{2.0cm}|>{\centering\arraybackslash}m{2.0cm}|}
\hline
\multirow{2}{*}{data set} & \multirow{2}{*}{$N_{\rm pts}$} & \multicolumn{2}{|c|}{$\Delta \chi^2$ relative to baseline} \\ \cline{3-4}
& & NNLO & aN${}^3$LO \\ \hline
NMC $\mu d$ $F_{2}$ \cite{NMC} &   123  &  $-4.2$  &  $-1.3$ \\
NuTeV $\nu N$ $F_{2}$ \cite{NuTev} &   53  & $-1.8$ & $-0.7$ \\
E866 / NuSea $pd/pp$ DY \cite{E866DYrat} &   15  & $-1.3$ & $-0.3$ \\
CCFR $\nu N \rightarrow \mu\mu X$ \cite{Dimuon} &   86  & $-2.3$ & $+0.4$ \\
NuTeV $\nu N \rightarrow \mu\mu X$ \cite{Dimuon} &   84  & $-2.2$ & $-0.9$ \\
HERA $e^{-}p$ CC \cite{H1+ZEUS} &   42  & $-1.6$ & $-0.0$ \\
HERA $e^{-}p$ NC $460\ \text{GeV}$ \cite{H1+ZEUS} &   209  &  $-1.7$ & $+0.6$ \\
HERA $e^{+}p$ NC $920\ \text{GeV}$ \cite{H1+ZEUS} &   402  & $-4.1$ & $+2.2$ \\
HERA $e^{-}p$ NC $575\ \text{GeV}$ \cite{H1+ZEUS} &   259  &  $-2.3$ & $+0.0$ \\
HERA $e^{-}p$ NC $920\ \text{GeV}$ \cite{H1+ZEUS} &   159  & $-1.2$ & $+1.1$ \\
D{\O} II $W \rightarrow \nu \mu$ asym. \cite{D0Wnumu} &   10  & $-1.5$ & $-0.4$ \\
LHCb $Z \rightarrow e^{+}e^{-}$ \cite{LHCb-Zee} &   9  &  $+0.9$ & $+0.5$ \\
LHCb W asym. $p_{T} > 20\ \text{GeV}$ \cite{LHCb-WZ} &   10  & $+1.2$ & $+0.4$ \\
ATLAS High-mass Drell-Yan \cite{ATLAShighmass} &   13  &  $+0.9$ & $+0.5$ \\
LHCb $8 \text{TeV}$ $Z \rightarrow ee$ \cite{LHCbZ8} &   17  & $+1.3$ & $+0.3$ \\
ATLAS $7\ \text{TeV}$ jets \cite{ATLAS:2014riz} & 140  &  $-3.4$ & $-0.6$ \\
CMS $7\ \text{TeV}\ W + c$ \cite{CMS7Wpc} &   10  & $+1.4$ & $-3.3$ \\
ATLAS $7\ \text{TeV}$ high prec. $W, Z$ \cite{ATLASWZ7f} &   61  & $-8.4$ & $-3.9$ \\
CMS $7\ \text{TeV}$ jets \cite{CMS:2014nvq} &   158  &  $-0.8$ & $-0.5$ \\
D{\O} $W$ asym. \cite{D0Wasym} &   14  & $-2.4$ & $-1.2$ \\
CMS $8\ \text{TeV}$ jets \cite{CMS:2016lna} &   174  & $+0.4$ & $-1.4$ \\
ATLAS $8\ \text{TeV}$ sing. diff. $t\bar{t}$ \cite{ATLASsdtop} &   25  & $+3.5$ & $+1.8$ \\
ATLAS $8\ \text{TeV}\ W + \text{jets}$ \cite{ATLASWjet} &   30  & $+1.2$ & $-1.1$ \\
CMS $8\ \text{TeV}$ double diff. $t\bar{t}$ \cite{CMS8ttDD} & 15  & $-1.4$ & $-1.3$ \\
ATLAS $8\ \text{TeV}\ W$ \cite{ATLASW8} &   22  & $-4.1$ & $-1.3$ \\
CMS $2.76\ \text{TeV}$ jet \cite{CMS276jets} &   81  & $+0.1$ & $-1.5$ \\
CMS $8\ \text{TeV}$ sing. diff. $t\bar{t}$ \cite{CMSttbar08_ytt} &   9  & $-2.0$ & $-0.7$ \\
ATLAS $8\ \text{TeV}$ double diff. $Z$ \cite{ATLAS8Z3D} &   59  & $-5.5$ & $-4.3$ \\
ATLAS $8\ \text{TeV}$ jets \cite{ATLAS:2017kux} & 171 & $-6.4$ & $-4.3$ \\ \hline
Total & 4430 & $-35.9$ & $-22.0$ \\ \hline
\end{tabular}
\vspace{0.1cm}
\caption {$\Delta \chi^2$ for the NNLO and aN${}^3$LO MSHT PDF fits upon removal of the ATLAS 8~TeV $Z$ $p_T$ data, negative indicates fit quality improvement upon its removal. This is only a selection of the data sets in the global fit, with those omitted which demonstrate little change.}
\label{tab:Zptremoval_chisqs}
\end{table}

In Table~\ref{tab:Zptcut_NNPDFCT_chisqs} we summarise the fit qualities obtained fitting different subsets of the data, including the exact data set fit by NNPDF and approximating the data set fit by CT. In the latter case for simplicity we still take the data as double differential in the $Z$ peak invariant mass bin for ease (resulting in 48 datapoints in total), whereas this is treated single differentially by CT. The uncertainty treatment remains as in MSHT20 in all cases, though as noted earlier this also varies between groups. A similar exercise was performed in the context of the MSHT20 NNLO PDF set in \cite{Bailey:2020ooq}, and we repeat it here in the context of an updated fit and extend it to aN${}^3$LO. We can then make a direct comparison of the aN${}^3$LO fit qualities obtained this way with like-for-like NNLO fit qualities. It can be seen that the fit quality at NNLO remains poor also for the data subset fit by NNPDF and the CT-like data subset. Some further improvement might be expected if the data in the $Z$ peak bin were rapidity integrated, which would reduce our CT-like fit from 48 to 27 datapoints (to match exactly the datapoints included in CT), as this places less constraints on the PDFs. Nonetheless, the fit qualities observed in Table~\ref{tab:Zptcut_NNPDFCT_chisqs} again suggest that the poor fit quality is not associated with the exact data points included at NNLO. At NNLO, CT and NNPDF quote fit qualities of $\sim 1.1$ \cite{Hou:2019efy} and $\sim 0.9$ \cite{NNPDF:2021njg} (also similar to that quoted in NNPDF3.1 \cite{NNPDF:2017mvq}) respectively and it was demonstrated in \cite{Bailey:2020ooq} that similar fit qualities could be obtained by adding an additional uncorrelated uncertainty of $1\%$ to the fit in the way is done in NNPDF for example. This suggests that differences seen between groups at NNLO reflect, rather than the data points included, instead other differences in the setups between the groups, including in the uncertainty treatment. In any case, returning to Table~\ref{tab:Zptcut_NNPDFCT_chisqs} we now observe that in all data selections the fit quality substantially improves at aN${}^3$LO, emphasising it is the higher order (aN${}^3$LO rather than NNLO) of the fit performed that leads to the possibility to fit these data well. 

\begin{table}
\fontsize{10}{12}\selectfont 
 \renewcommand\arraystretch{1.05} 
\centering \hspace{-0.1cm} \begin{tabular}{|>{\centering\arraybackslash}m{2.25cm}|>{\centering\arraybackslash}m{2.75cm}|>{\centering\arraybackslash}m{2.75cm}|>{\centering\arraybackslash}m{2.25cm}|} 
\hline
Fit Order & MSHT Default & NNPDF-like & CT-like \\ \hline
NNLO & 1.87 & 1.80 & 1.79 \\  
aN${}^3$LO & 1.04 & 1.02 & 0.87 \\ \hline
$N_{\rm pts}$ & 104 & 92 & 48 \\ \hline 
\end{tabular}
\vspace{0.1cm}
\caption {Fit qualities, i.e. $\chi^2/N_{\rm pts}$, for NNLO and aN${}^3$LO MSHT PDF fits, varying the data cut to match the NNPDF included data (NNPDF-like) or to approximate the CT included data (CT-like).}
\label{tab:Zptcut_NNPDFCT_chisqs}
\end{table}

We can further examine the nature of the better fit quality at aN$^{3}$LO to the ATLAS 8~TeV $Z$ $p_T$ data set by making further intermediate PDF fits, with certain aspects of the aN$^{3}$LO theory incorporated or not and then comparing the fit qualities of these intermediate fits with the full NNLO or aN$^{3}$LO. One particular question to address, is the extent to which the fit quality improvement is associated to the genuine and largely known included aN$^{3}$LO effects in the splitting functions (as well as indirectly from the known N$^{3}$LO ingredients included for the transition matrix elements and DIS coefficient functions - for more information see \cite{McGowan:2022nag}) or the K-factor freedom allowed for the unknown N$^{3}$LO vector boson plus jets K-factor, which contains theoretical nuisance parameters in the aN$^{3}$LO fit. The details of the procedure for the aN$^{3}$LO (fitted) K-factors included in the aN$^{3}$LO PDFs and their uncertainties are given in Section~7 of \cite{McGowan:2022nag}.
It was observed in \cite{McGowan:2022nag}, see e.g. Table~4, as well as earlier in this work in the context of the jets and dijet analysis (see Table~\ref{tab:jets_chi2} of Section~\ref{sec:fitquality}), that the fit to the $Z$ $p_T$ data at aN$^{3}$LO even with NNLO K-factors (the final columns) is notably better, indicating a preference not only for the shape and normalisation freedom provided by our fitted aN$^{3}$LO K-factors (see Section 7 of \cite{McGowan:2022nag}) but also for the known N$^{3}$LO information encoded in the higher order splitting functions, DIS coefficient functions and transition matrix elements of the aN$^{3}$LO PDF fit. We examine this in greater detail here.

In Table~\ref{tab:Zpt_kfs_fitqualities} the fit qualities of the ATLAS 8~TeV $Z$ $p_T$ data in a full NNLO global PDF fit (second column) and the full aN$^{3}$LO global PDF fit (last column) are therefore compared with two intermediate PDF fits; in the fourth column with an aN$^{3}$LO fit with the purely known NNLO K-factors (as provided previously), whilst the third column provides the ``inverse case'' of an NNLO fit with additional aN$^{3}$LO K-factors (determined by the fit). The total fit qualities of these fits are given in the bottom row. This comparison demonstrates that  the aN$^{3}$LO K-factor freedom and the remainder of the largely known aN$^{3}$LO ingredients  both contribute relatively equally to the fit quality improvement. The NNLO fit has a relatively poor fit quality of 1.87 per point, the two intermediate cases both have very similar and notably improved fit qualities of $\approx 1.4$ per point, whilst the full aN$^{3}$LO fit improves to 1.04 per point. This demonstrates that whilst the aN$^{3}$LO K-factor freedom is able to account for some of the data and theory differences which result in a poor fit at NNLO, it is not able to account for much of the difference due presumably to non-trivial shape or normalisation differences. These cannot be absorbed into the unknown N3LO K-factors and so the fit quality is only improved further by the inclusion of the other largely known aN$^{3}$LO theory ingredients, as one would hope.  Corresponding improvements in the global PDF fit quality are also seen, as provided in the bottom row of Table~\ref{tab:Zpt_kfs_fitqualities}.

\begin{table}
\fontsize{10}{12}\selectfont 
 \renewcommand\arraystretch{1.05} 
\centering \hspace{-0.1cm} \begin{tabular}{|>{\centering\arraybackslash}m{4.55cm}|>{\centering\arraybackslash}m{1.5cm}|>{\centering\arraybackslash}m{2.65cm}|>{\centering\arraybackslash}m{2.65cm}|>{\centering\arraybackslash}m{1.5cm}|} 
\hline 
$\chi^2/N_{\rm pts}$ & NNLO & NNLO + aN$^{3}$LO K-factors free & aN$^{3}$LO (NNLO K-factors) & aN$^{3}$LO \\ \hline
ATLAS 8~TeV $Z p_T$ & 1.87 & 1.41 & 1.37 & 1.04 \\ \hline
Total & 1.22 & 1.19 & 1.19 & 1.17 \\ \hline
\end{tabular}
\vspace{0.1cm}
\caption {Fit qualities, i.e. $\chi^2/N_{\rm pts}$, for in turn our default NNLO fit, an NNLO fit with fitted aN$^{3}$LO K-factors used, an aN$^{3}$LO fit with the NNLO K-factors used and finally our default aN$^{3}$LO fit. The fit qualities of the ATLAS 8~TeV $Z p_T$ data and the total global fit are shown.}
\label{tab:Zpt_kfs_fitqualities}
\end{table}

\begin{figure}
\begin{center}
\includegraphics[scale=0.22,trim=0.2cm 0cm 0.1cm 0cm,clip]{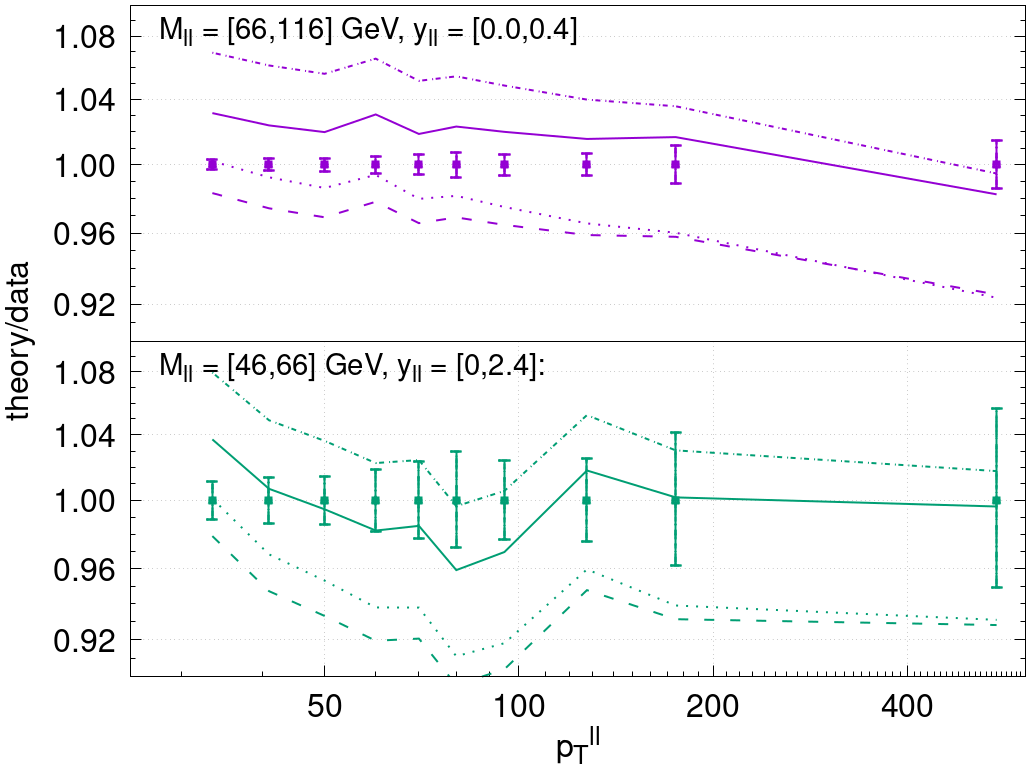} 
\includegraphics[scale=0.22,trim=1.25cm 0cm 0.1cm 0cm,clip]
{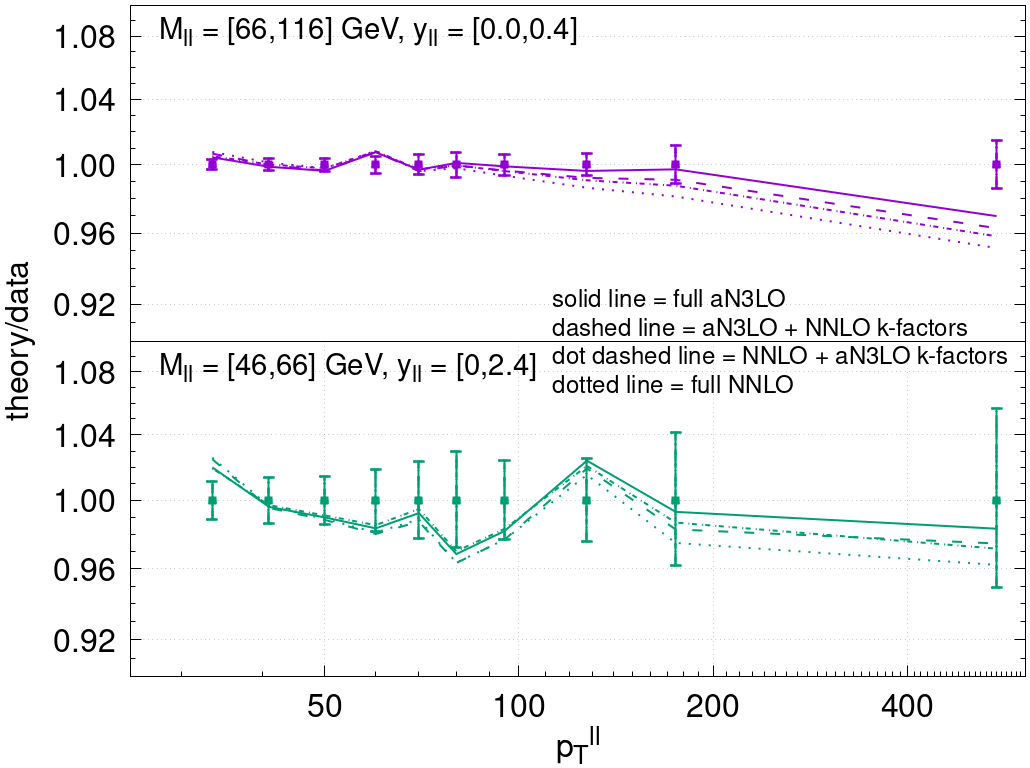} 
\caption{\sf Theory and data comparison at NNLO (dotted lines), aN$^{3}$LO but with NNLO K-factors only (dashed lines), NNLO with aN$^{3}$LO fixed K-factors added (fine dotted lines), NNLO with aN$^{3}$LO free K-factors added (dot dashed lines) and aN$^{3}$LO (solid lines). The ratio of theory/data is shown and the points are the datapoints with their total uncorrelated error only shown by the errorbars. The left figure is the data pre-fit, whilst the right figure is after the correlated systematic pulls are accounted for. Representative bins are shown in each figure, with the top being the $M_{ll}=[66,116]~{\rm GeV}, y_{ll}=[0.0,0.4]$ and the bottom being the $M_{ll}=[46,66]~{\rm GeV}$ $y_{ll}=[0.0,2.4]$ bin. }
\label{fig:ZpT_datatheory_unandshifted_twoonly}
\end{center}
\end{figure}

This observation is further elucidated by the direct comparisons of the theoretical predictions obtained in the various PDF fits with the data. Ratios of theory to data, both for the raw pre-fit data unaltered by the correlated systematics and with the data post-fit after these are taken into account by the fit, are shown for two representative cases are shown in Fig.~\ref{fig:ZpT_datatheory_unandshifted_twoonly}. Specifically the two cases shown are for the $Z$ peak $M_{ll}=[66,116]~{\rm GeV}$, $y_{ll}=[0.0,0.4]$ bin and the $M_{ll}=[46,66]~{\rm GeV}$ rapidity integrated ($y_{ll}=[0.0,2.4]$) bin. The left figure displays the comparison for the raw pre-fit data, and the right figure after allowing the correlated systematic shifts. The same four PDF fits considered in Table~\ref{tab:Zpt_kfs_fitqualities} are shown. The comparison plots for all of the six double-differential rapidity bins in the Z peak mass bin of [66,116]~GeV and all of the five single-differential other mass bins, before and after the correlated systematic pulls are applied in the fit, are given in Appendix~\ref{app:Zpt_datatheorycomp}. The solid line is the full aN$^{3}$LO theory/data ratio, for comparison against the datapoints with their uncorrelated uncertainty shown by their errorbars. In the pre-fit case, it is visible by eye in several of the bins that this is more closely following the datapoints, though differences still remain. The full NNLO (dotted line), NNLO + aN$^{3}$LO K-factor case (dot dashed line), and aN$^{3}$LO + NNLO fixed K-factors (dashed line) are not able to follow the data to the same degree. After accounting for the correlated systematic pulls in the fit, all four cases naturally move closer to the datapoints, as must occur, though it is still visible even by eye that the full aN$^{3}$LO appears better able to describe the datapoints in several of the bins. This is reflected in the fit qualities given in Table~\ref{tab:Zpt_kfs_fitqualities}. This further highlights the need for the aN$^{3}$LO theory to describe these precise data. 

This is also reflected by the average theory/data ratio across all 104 datapoints, which whilst a relatively crude measure, is also closer to 1 in the full aN$^{3}$LO case both for the pre-fit data and after shifting by the correlated systematics in the fit than for any of the other 3 fits shown in the figure. 
As a result of the better agreement of the aN$^{3}$LO fit, even for the pre-fit, it also has the smallest contribution to the $\chi^2$ from the correlated systematic experimental nuisance parameter penalties, whereas the NNLO fit has the largest. The penalties from the correlated systematic pulls at NNLO account for $+37.0$ of the $\Delta\chi^2 = +85.4$ deterioration in the fit quality at NNLO relative to aN$^{3}$LO.  In the two intermediate cases (NNLO + aN$^{3}$LO K-factors or aN$^{3}$LO + NNLO K-factors) where we observe approximately half the improvement in fit quality, as seen previously in Table~\ref{tab:Zpt_kfs_fitqualities}, the penalties from the correlated systematic pulls are also approximately halved relative to the NNLO case. This indicates that the pulls of these in the lower order (not full aN$^{3}$LO) cases attempt to absorb some of the theory - data difference, which is substantially reduced by the aN$^{3}$LO theory.
All in all, these observations demonstrate the need for the aN$^{3}$LO theory to describe these $Z$ $p_T$ data, with a preference for not simply the freedom associated with our treatment of the unknown aN$^{3}$LO K-factors, but rather the full aN$^{3}$LO fit. This indicates that the inclusion of the known aN$^{3}$LO information in the full global PDF fit is required to obtain a good fit of these data.

\subsection{Impact on PDFs}\label{sec:zptPDFs}

\begin{figure}[t]
\begin{center}
\includegraphics[scale=0.21]{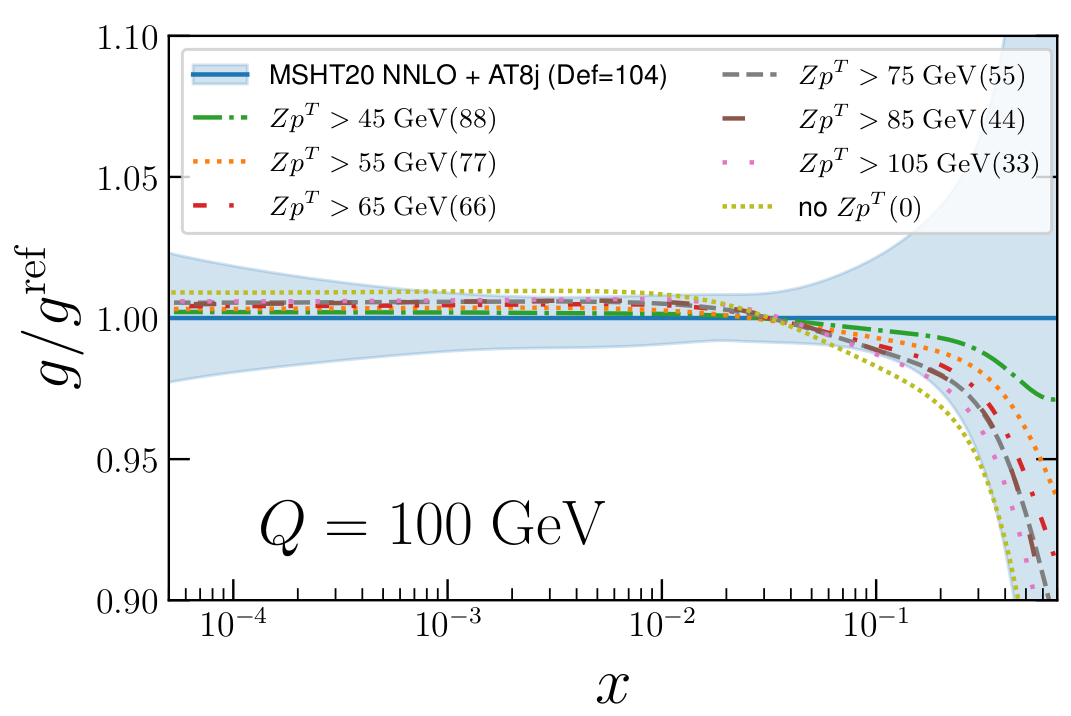}
\includegraphics[scale=0.21,trim=3.25cm 0cm 0.1cm 0cm,clip]{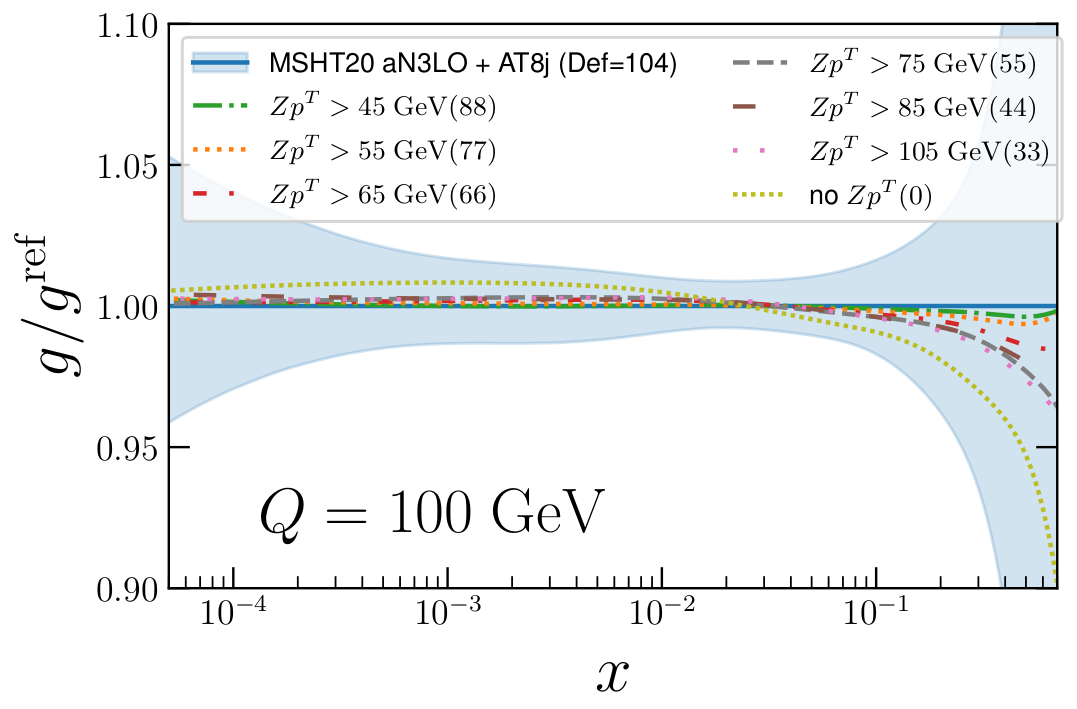}
\caption{\sf The effect on the gluon PDF of cutting data below a given $p_T^{ll}$ from the ATLAS 8~TeV $Z$ $p_T$ data at NNLO (left) and aN${}^3$LO (right) at $Q=100~{\rm GeV}$. The numbers in brackets give the number of datapoints included.}
\label{fig:gluon_Zptcuts}
\end{center}
\end{figure}

It is also instructive to examine the effects of these $Z$ $p_T$ data on the PDFs, in particular the high $x$ gluon PDF, on which it places constraints. It was demonstrated that for NNLO PDFs these data resulted in a net upward pull on the gluon at high $x$ in \cite{Bailey:2020ooq}, and similar effects are seen at aN${}^3$LO (this is also shown later in Fig.~\ref{fig:gluon_Zptcuts}). Ultimately this has a significant impact on the resulting gluon at high $x$. This is also noted at NNLO albeit to a lesser degree by NNPDF4.0 \cite{NNPDF:2021njg}, and CT18 \cite{Hou:2019efy} see smaller effects still - though both groups include fewer datapoints than in MSHT and there are additional differences in their treatments of the uncertainties, as discussed in the previous section. The pulls of these data on the MSHT20 PDF fits were also shown in \cite{Jing:2023isu}, though in this section we will examine this in more detail, in the context of the various fits with different data cuts and selections described in the previous section (Section~\ref{sec:zptfitquality}).

\begin{figure}[t]
\begin{center}
\includegraphics[scale=0.2]{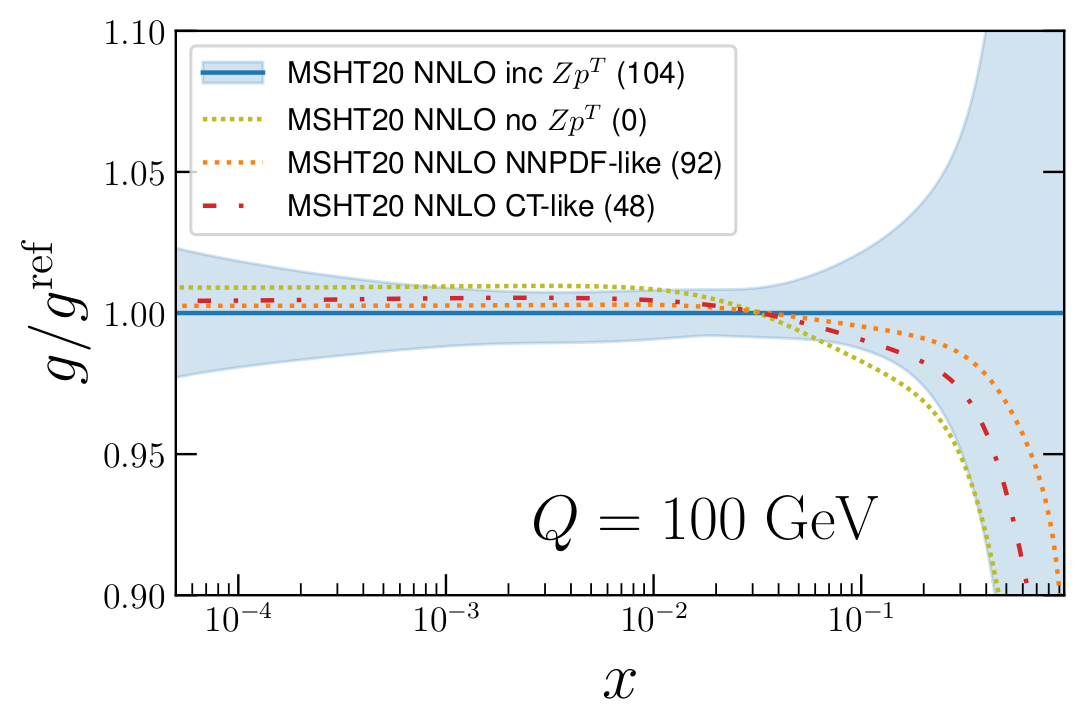} 
\caption{\sf The effect on the gluon PDF at NNLO of varying the $Z$ $p_T$ datapoints included to match those included by NNPDF (NNPDF-like) or to approximate those included by CT (CT-like), more details are given in the text. The numbers in brackets give the number of datapoints included.}
\label{fig:gluon_NNLO_ZpT_CTNNPDFlike}
\end{center}
\end{figure}

\begin{figure}
\begin{center}
\includegraphics[scale=0.205]{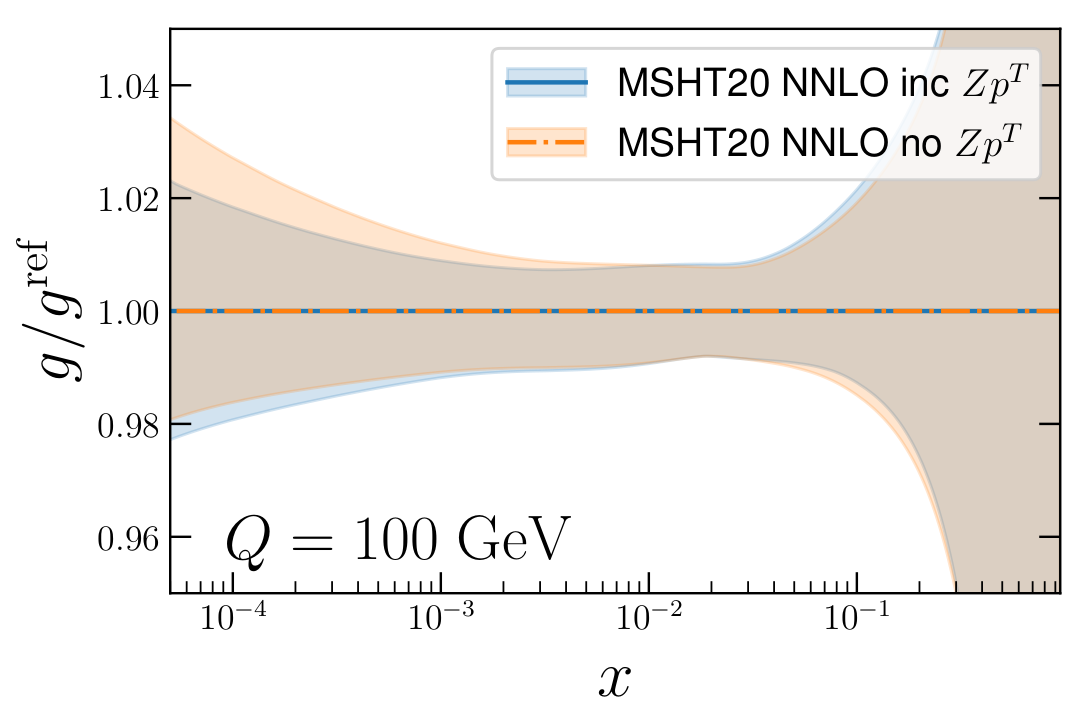}
\includegraphics[scale=0.205]{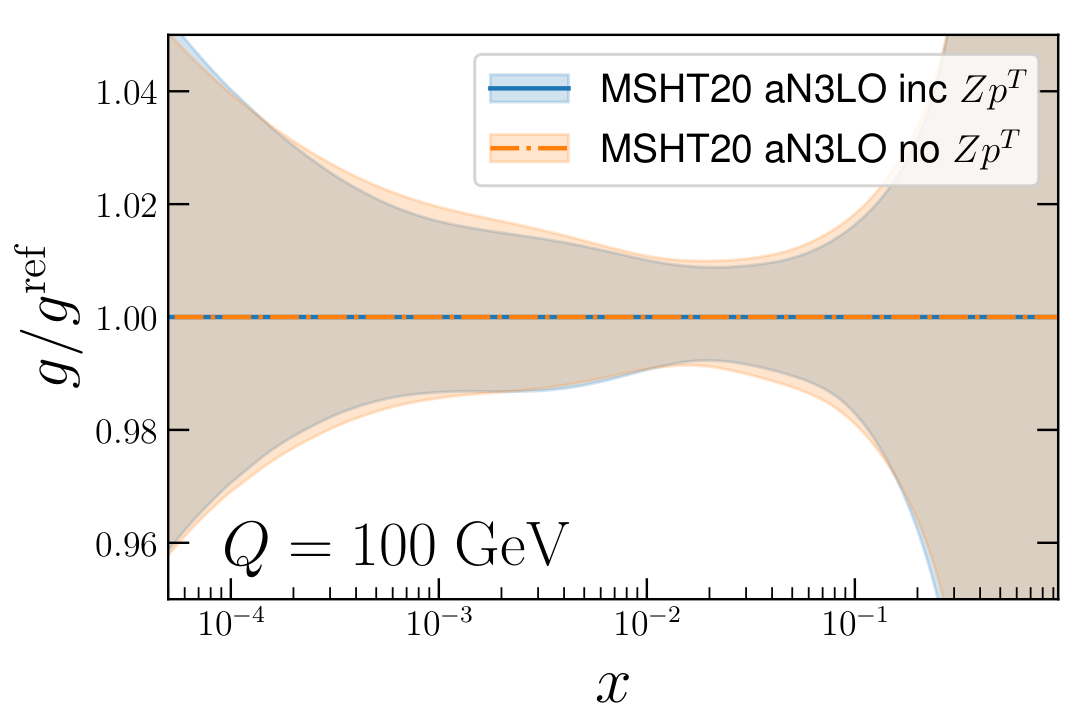}
\caption{\sf The effect on the gluon PDF uncertainty of removing the ATLAS 8~TeV $Zp_T$ data set at NNLO (left) and aN${}^3$LO (right) at $Q=100~{\rm GeV}$.}
\label{fig:gluon_Zpterrs}
\end{center}
\end{figure}

\begin{figure}
\begin{center}
\includegraphics[scale=0.19, trim=0 -1cm 0 0, clip]{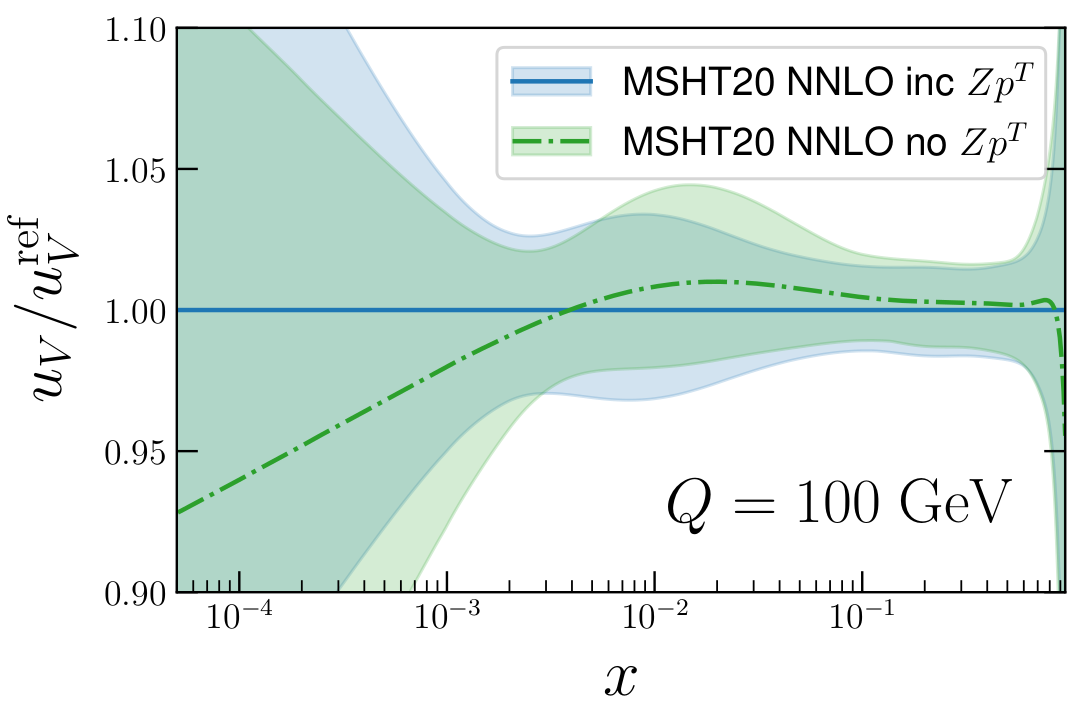}
\includegraphics[scale=0.19, trim=0 -1cm 0 0, clip]{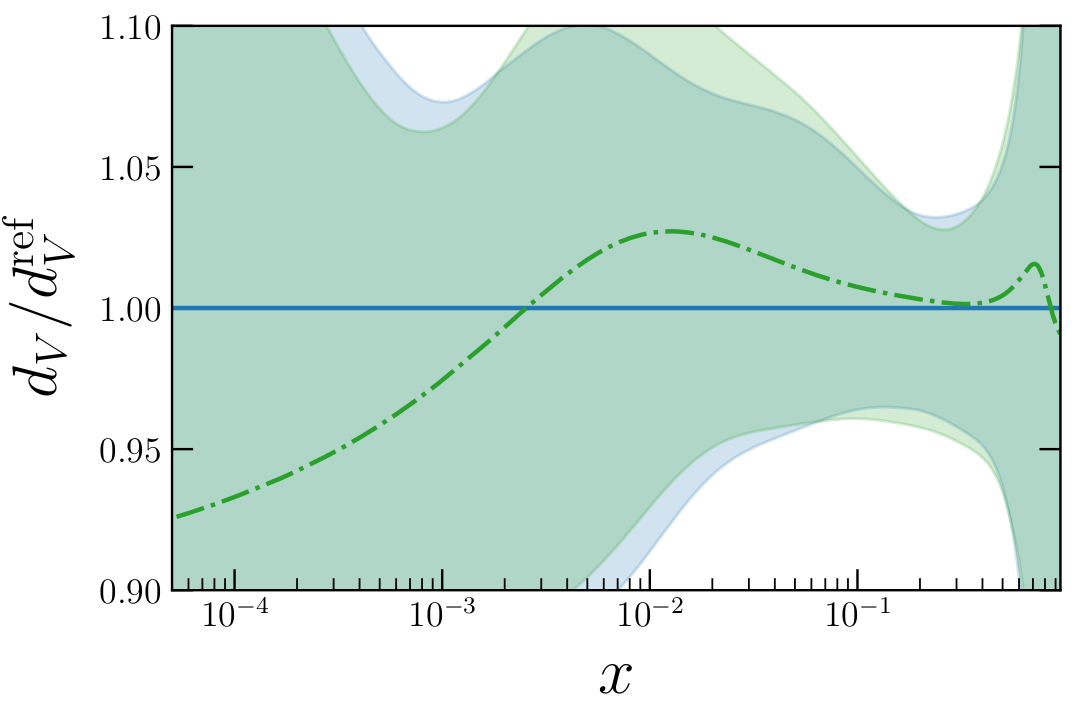}
\includegraphics[scale=0.19, trim=0 -1cm 0 0, clip]{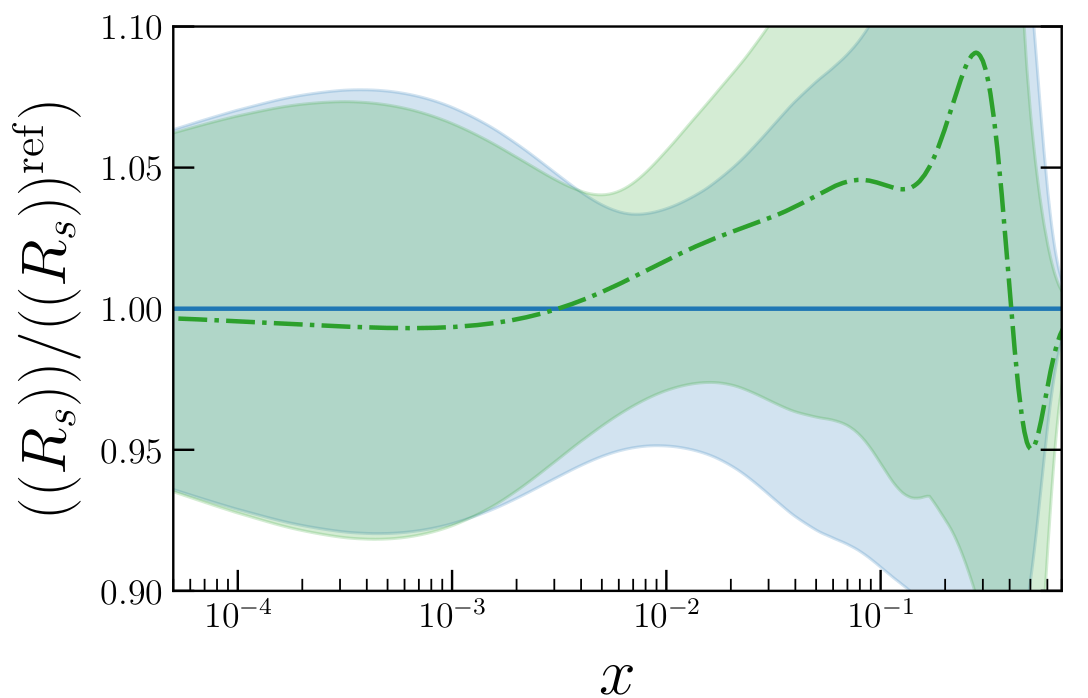}
\includegraphics[scale=0.19, trim=0 -1cm 0 0, clip]{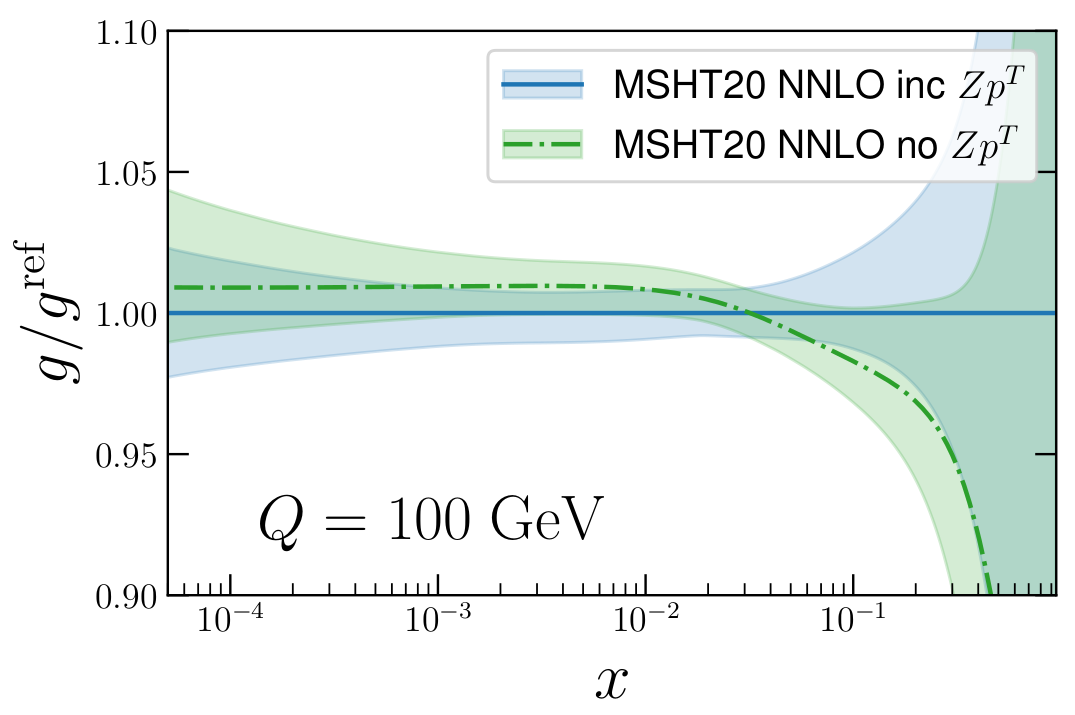}
\caption{\sf The effect of the ATLAS 8~TeV $Zp_T$ data set at NNLO on the (upper left) up valence, (upper right) down valence, (lower left) strangeness ratio $R_s = \frac{s+\bar{s}}{\bar{u}+\bar{d}}$, and (lower right) gluon PDFs at $Q=100~{\rm GeV}$.}
\label{fig:ZpT_PDFs_NNLO}
\end{center}
\end{figure}

\begin{figure}
\begin{center}
\includegraphics[scale=0.19, trim=0 -1cm 0 0, clip]{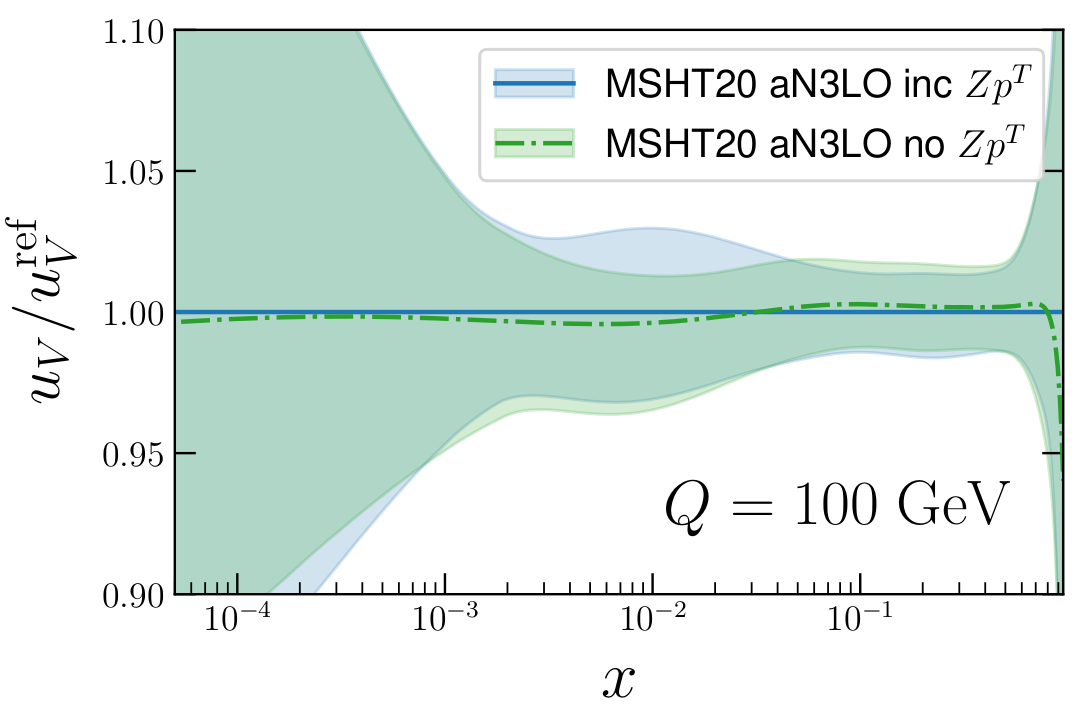}
\includegraphics[scale=0.19, trim=0 -1cm 0 0, clip]{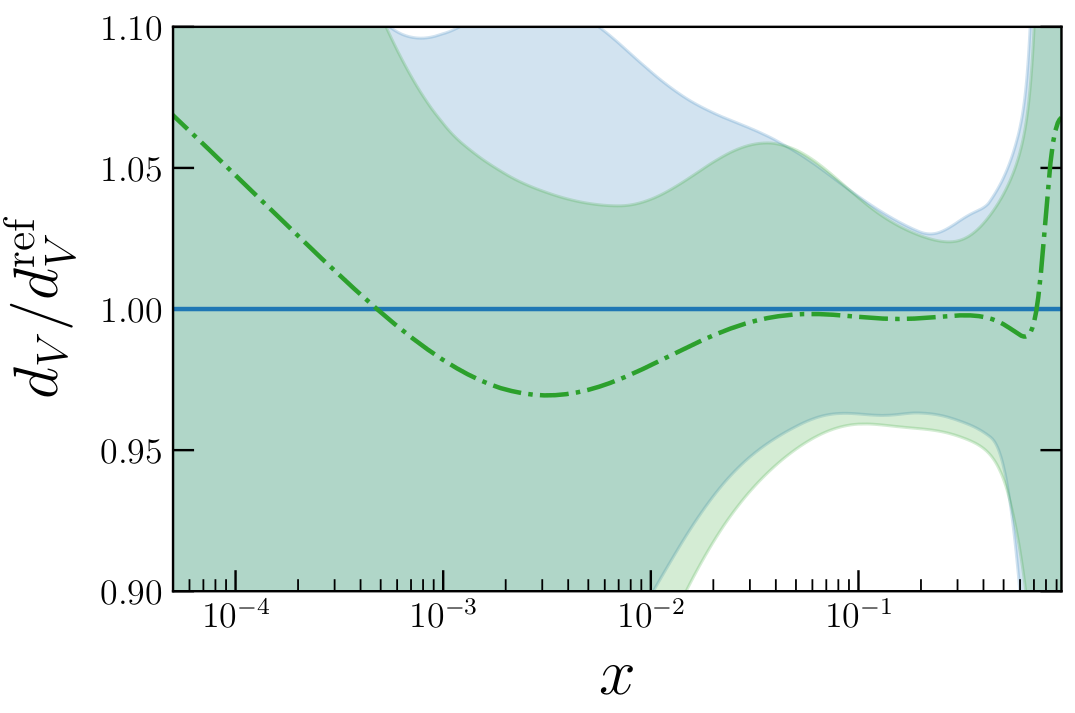}
\includegraphics[scale=0.19, trim=0 -1cm 0 0, clip]{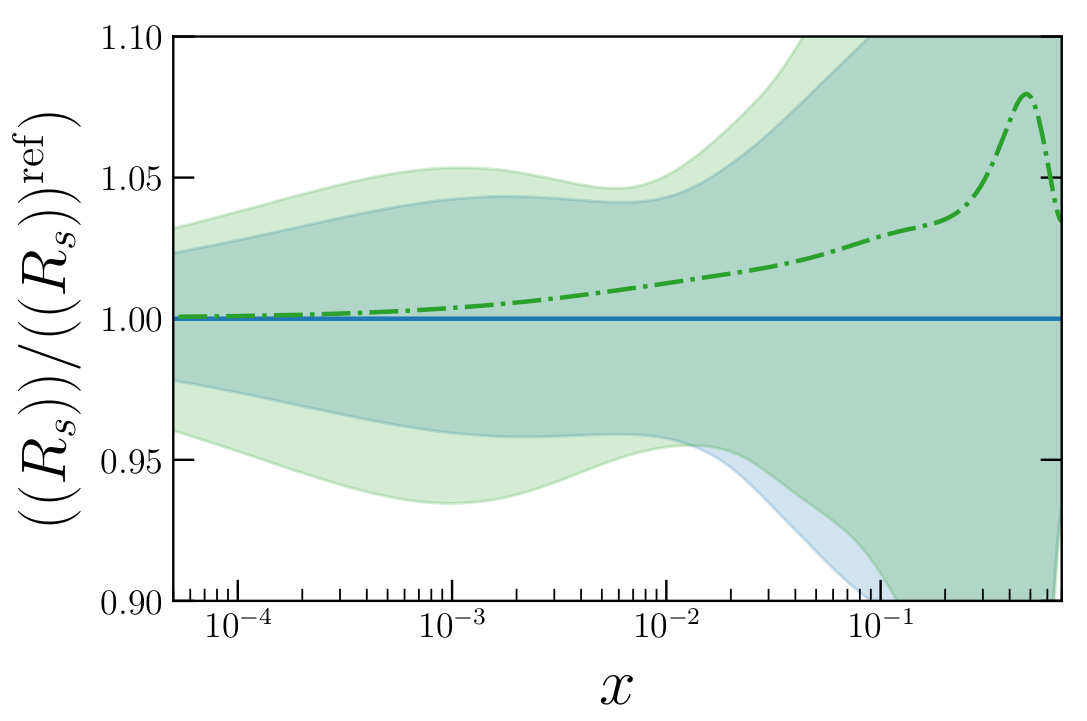}
\includegraphics[scale=0.19, trim=0 -1cm 0 0, clip]{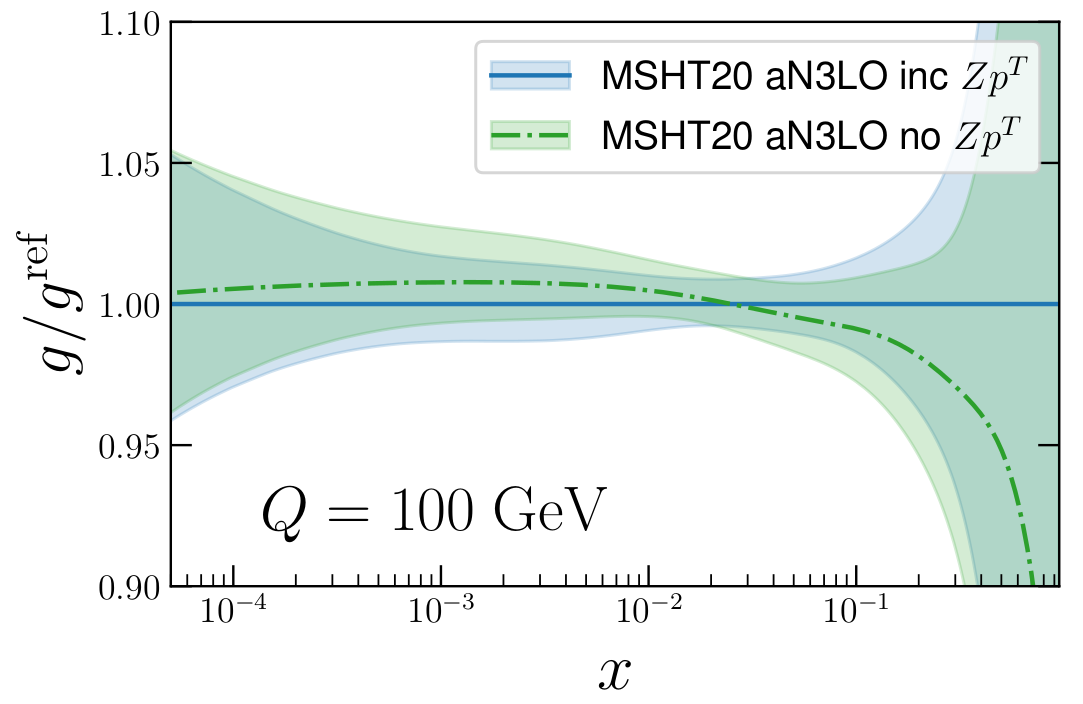}
\caption{\sf The effect of the ATLAS 8~TeV $Zp_T$ data set at aN${}^3$LO on the (upper left) Up valence, (upper right) Down valence, (lower left) Strangeness ratio $R_s = \frac{s+\bar{s}}{\bar{u}+\bar{d}}$, and (lower right) gluon PDFs at $Q=100~{\rm GeV}$.}
\label{fig:ZpT_PDFs_N3LO}
\end{center}
\end{figure}

In Fig.~\ref{fig:gluon_Zptcuts} the effect of these $Z$ $p_T$ data on the high $x$ gluon PDF at NNLO (left figure) and aN${}^3$LO (right figure) are shown. The denominators of the ratios are the new PDFs at their respective orders with the ATLAS 8~TeV inclusive jet data included on top of the MSHT20 PDFs, with the uncertainty band on this new baseline also shown. The dotted light green lines indicate the impact of removing this ATLAS 8~TeV $Z$ $p_T$ data set completely is a reduction in the high $x$ gluon for $x \gtrsim 0.035$. It is immediately clear that the impact of the removal of the data set at aN${}^3$LO is smaller than at NNLO, with the latter producing a gluon around the edge of (or in places beyond) the uncertainty band whilst the former lies well inside the gluon PDF uncertainty. Moreover, gradually cutting out more of the data by increasing the $p_T^{ll}$ cut smoothly transitions from the default fit towards the case with no $Z$ $p_T$ data, again reinforcing the fact that no issues with particular $p_T^{ll}$ regions are seen. In the context of our investigations mimicking the NNPDF and CT datapoint selections (``CT-like'' and ``NNPDF-like'') in the previous Section~\ref{sec:zptfitquality}, we also observe that the CT-like and NNPDF-like fits show similar pulls on the high $x$ gluon PDF, though analogously reduced in magnitude due to the fewer number of datapoints included. This is shown in Fig.~\ref{fig:gluon_NNLO_ZpT_CTNNPDFlike} at NNLO, a similar effect is seen at aN${}^3$LO and so is not shown here, though with slightly reduced pulls.

Fig.~\ref{fig:gluon_Zpterrs} illustrates the impact of these data on the gluon PDF uncertainty. At NNLO, the impact of the presence or absence of the ATLAS 8~TeV $Z$ $p_T$ data set on the high $x$ gluon uncertainty is very small, despite the notable constraints it offers on the central value, as a result of the significant tensions it shows with other data in the global fit. In contrast, at aN${}^3$LO there is a somewhat clearer reduction on the gluon uncertainty in the $x\sim 0.05$ region from the inclusion of the ATLAS 8~TeV $Z$ $p_T$ data set. The reduced tensions between it and other data at aN${}^3$LO enable it to further constrain the gluon PDF uncertainty.

Finally, in Figs.~\ref{fig:ZpT_PDFs_NNLO} and \ref{fig:ZpT_PDFs_N3LO} the effect of the inclusion of the ATLAS 8~TeV $Z$ $p_T$ data set on several other PDFs is shown, including the gluon but now also the up and down valence quarks and the strangeness ratio $R_s = \frac{s+\bar{s}}{\bar{u}+\bar{d}}$. In addition to the pull on the high $x$ gluon, the $Z$ $p_T$ data also have a pull on the quark PDFs, as these initiate the production of a $Z$ boson at leading order. The quark PDFs are relatively well constrained and so the impact is smaller but there are visible pulls downwards on the up and down valence PDFs and the strangeness ratio at $x \gtrsim 0.01$ at NNLO. On the other hand, the effect is again milder at aN${}^3$LO, perhaps due to the reduced tension in this fit. The strangeness ratio nonetheless remains lowered once the $Z$ $p_T$ data is included at intermediate to large $x$. This may be due to the momentum sum rule. The $Z$ $p_T$ data prefers a gluon which is larger at higher $x$ values, and hence carries more momentum. This must come from somewhere, and the high-$x$ strange quark is the least well-constrained PDF carrying appreciable momentum.

%\newpage

\section{Conclusions}\label{sec:conc}

In this paper we have presented analyses of two key sets of data for constraining the high $x$ gluon, comparing their impacts at both NNLO and a${\rm N}^3$LO. 

First we have presented the first analysis of LHC inclusive jet and dijet production in a global PDF fit at up to a${\rm N}^3$LO order. We have analysed in detail the fit quality, consistency between the jet and dijet cases, and overall PDF impact of these data sets. We have observed that at NNLO a good fit quality to LHC 7 and 8 TeV data on dijet production can be achieved, in contrast to the case of inclusive jet production, where the fit quality is rather poor, consistent with earlier studies. This remains true at a${\rm N}^3$LO, i.e. the fit to the inclusive jet data continues to be poor. In addition, we have observed that the inclusion of EW corrections in fact leads to some deterioration in the fit quality at either QCD order in the inclusive jet case, in contrast to what we might expect. For dijet production, on the other hand, the fit quality does improve upon the inclusion of EW corrections. The above results are found to be stable under a change of the choice of scale for the inclusive jet case, namely between $p_\perp^j$ and $H_\perp$; some improvement is observed when using the former scale at NLO, but at NNLO order and higher the fit quality is rather stable. At NNLO, we have also found evidence for a reduced tension between the dijet data and the ATLAS 8 TeV $Z$ $p_T$ data, which is also sensitive to the high $x$ gluon, although at a${\rm N}^3$LO this tension is largely eliminated in both the jet and dijet fits. 

We have also observed that at NLO the fit quality to the CMS 8 TeV triple differential dijet data (the only data of this type so far included in a PDF fit) is extremely poor; it is only by going to at least NNLO that a good fit can be achieved. This highlights the importance of having high precision theoretical calculations to match such high precision and multi--differential data.

In terms of the impact on the gluon PDF, we note some moderate difference in pull between the inclusive jet and dijet fits at NNLO, although these are statistically compatible. Therefore at this order the choice of which data set to include in the fit will have some effect on the extracted PDFs. On the other hand, at a${\rm N}^3$LO this difference is largely eliminated and the resulting PDFs are rather compatible, in particular in the case of the gluon. A further interesting observable to consider is the impact on the fit value of the strong coupling, $\alpha_S$, which will be the topic of a future study. 

A further issue that we have made the first  study of within the context of a global PDF fit is that of full colour corrections at NNLO in the theoretical predictions. We have studied the impact of these when they are included for the most relevant data set, namely the CMS 8 TeV triple--differential dijet data. We find that at NNLO there is in fact a relatively mild deterioration in the fit quality, but which is not present at  a${\rm N}^3$LO. The impact on the gluon PDF is found to be mild, but not completely negligible. Therefore, in general the impact of FC corrections is found to be relatively mild, but not necessarily insignificant. Until these are available for the full range of jet and dijet data considered it is arguably difficult to draw completely firm conclusions about this.

From these results, we  conclude that at NNLO it may be preferable to include dijet rather than inclusive jet production data in future PDF fits, when a choice must be made between the two, as is currently the case for all existing jet and dijet measurements that derive from the same data set. This choice is found to have some impact on the PDFs at NNLO, and it will therefore be of great interest for experimental collaborations to provide the full correlations between the jet and dijet data sets, such that a simultaneous analysis can be performed. Absent this, the increased stability of the PDFs with respect to this choice at a${\rm N}^3$LO  is further evidence that going to this order (and/or including missing higher order uncertainties in the cross section calculations) may be preferable. The above conclusions are found to be independent of whether FC corrections are applied at NNLO for the most sensitive CMS 8 TeV dijet data set.

We have in addition revisited the analysis of the ATLAS 8 TeV $Z$ $p_T$ data. At NNLO, we find the fit quality to this data set remains poor, regardless of the particular cuts and datapoint selections made. The fit quality at aN${}^3$LO is not only always improved relative to NNLO but also well fit regardless of the datapoints included. We therefore conclude that we find no evidence that the poor fit quality at NNLO is related to issues with particular parts of the $p_T^{ll}$ spectrum but rather the general need to go to higher orders in the fit. Moreover, we demonstrate that the improvement in fit quality is associated not only with the additional freedom provided by the unknown a${\rm N}^3$LO K-factors, but also to a significant extent with the known a${\rm N}^3$LO information included in the fit. 
The impact on the fit of these ATLAS 8~TeV $Z$ $p_T$ data is consistent between NNLO and aN${}^3$LO, both resulting in an upward pull on the gluon at high $x$, though the pull is somewhat reduced at aN${}^3$LO.  Greater tensions between the $Z$ $p_T$ data and other data in the global fit are also seen at NNLO than at aN${}^3$LO. As a result, these restrict the ability of the $Z$ $p_T$ data to further constrain the gluon PDF uncertainty at high $x$ at NNLO to the extent seen at  aN${}^3$LO.  Overall we therefore conclude that the improvement of the fit quality and reduction in tensions of these precise $Z$ $p_T$ data at aN${}^3$LO relative to NNLO is a sign of a genuine preference for higher order QCD effects. 

In summary, both the studies of jet and dijet data and the ATLAS 8~TeV $Z$ $p_T$ data have demonstrated the need for precise theory to match experimental precision. The aN${}^3$LO theoretical accuracy provided by the MSHT PDFs is found in both cases to be a key ingredient in this.

\section*{Acknowledgements}

We thank Jamie McGowan, whose invaluable work on the original a${\rm N}^3$LO fit provided the groundwork for this study. We thank the NNPDF collaboration and Emmanuele Nocera in particular for providing NLO theory grids for certain jet and dijet data sets. We thank Alex Huss for providing NNLO K-factors for the jet and dijet data. We thank Engin Eren and Katerina Lipka for help with CMS data, in particular the statistical correlations for the 8 TeV inclusive jet data. We thank Klaus Rabbertz for helpful correspondance about the experimental error definitions, and their signs, in the CMS 8 TeV dijet data.
We also thank our co-authors on ref. \cite{Jing:2023isu} for discussions of the $Z$ $p_T$ pulls. TC acknowledges that this project has received funding from the European Research Council (ERC) under the European Union’s Horizon 2020 research and innovation programme (Grant agreement No. 101002090 COLORFREE).  L. H.-L. and R.S.T. thank STFC for support via grant awards ST/T000856/1 and ST/X000516/1.  

\appendix
\section{ATLAS 8~TeV $Z$ $p_T$ Theory-Data Comparison}\label{app:Zpt_datatheorycomp}

Here we provide theory data comparison plots for all bins, including all 104 datapoints, for the ATLAS 8~TeV $Z$ $p_T$ data set, first for the pre-fit data before accounting for the correlated systematic pulls and then for the data shifted by the fits by the correlated systematics.

\begin{figure}
\begin{center}
\includegraphics[height=16cm, width=16cm,trim=0.2cm 0cm 0.1cm 0cm,clip]
{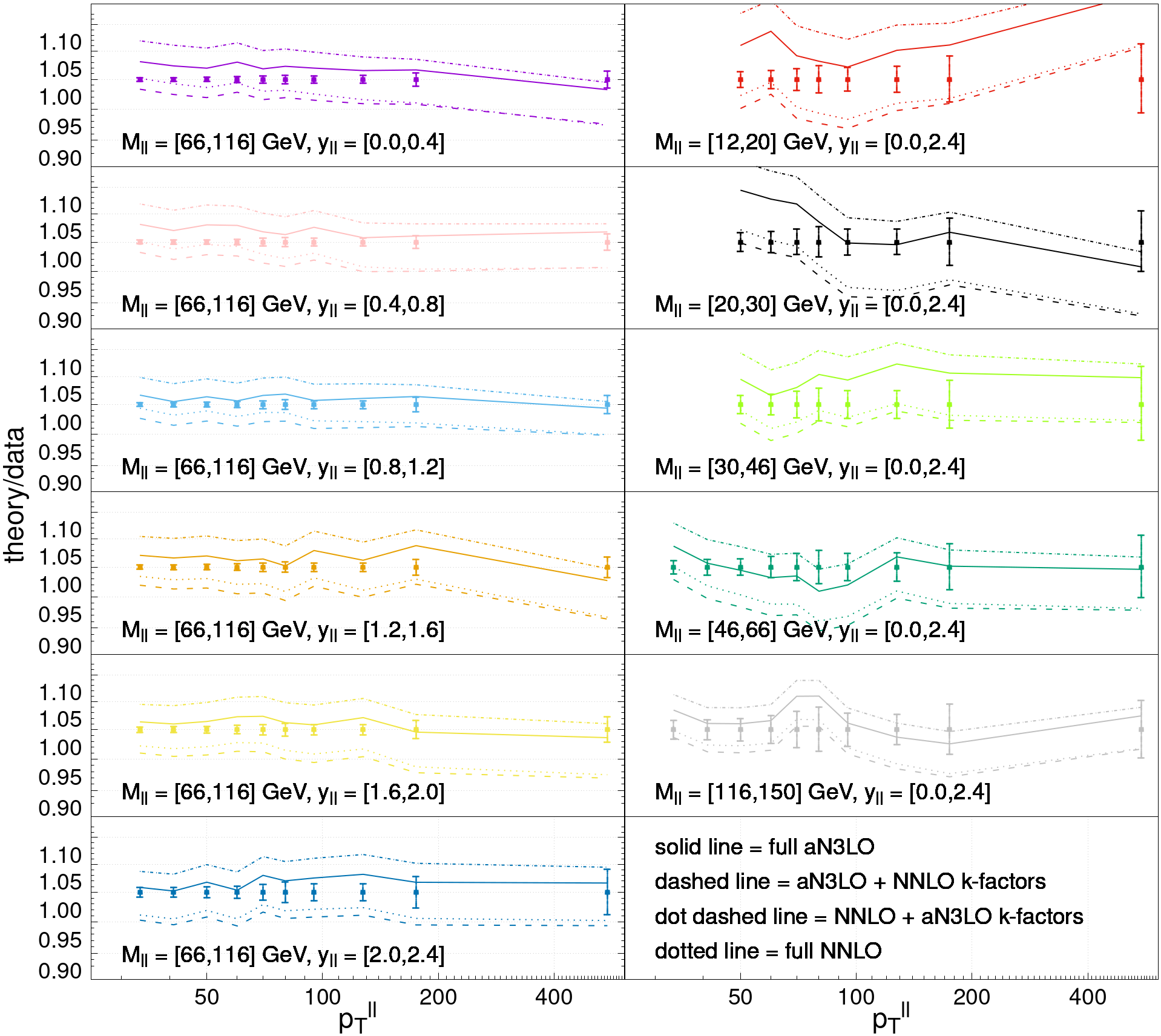}
\caption{\sf Theory and data comparison at NNLO (dotted lines), NNLO with aN$^{3}$LO K-factors added (dot dashed lines), aN$^{3}$LO with NNLO K-factors (dashed lines) and aN$^{3}$LO (solid lines). The ratio of theory/data is shown and the points are the datapoints with their total uncorrelated error (quadrature sum of their statistical and uncorrelated systematic errors) only shown by the errorbars. On the left are the 6 double differential bins in $p_T^{ll}$ and $y_{ll}$ the $Z$ peak region, and on the right are the 5 single differntial bins in $p_T^{ll}$ for the other mass bins. In this figure the data are before shifting by the fit of the correlated systematic pulls.}
\label{fig:ZpT_datatheory_unshifted}
\end{center}
\end{figure}

\begin{figure}
\begin{center}
\includegraphics[height=16cm, width=16cm,trim=0.2cm 0cm 0.1cm 0cm,clip]{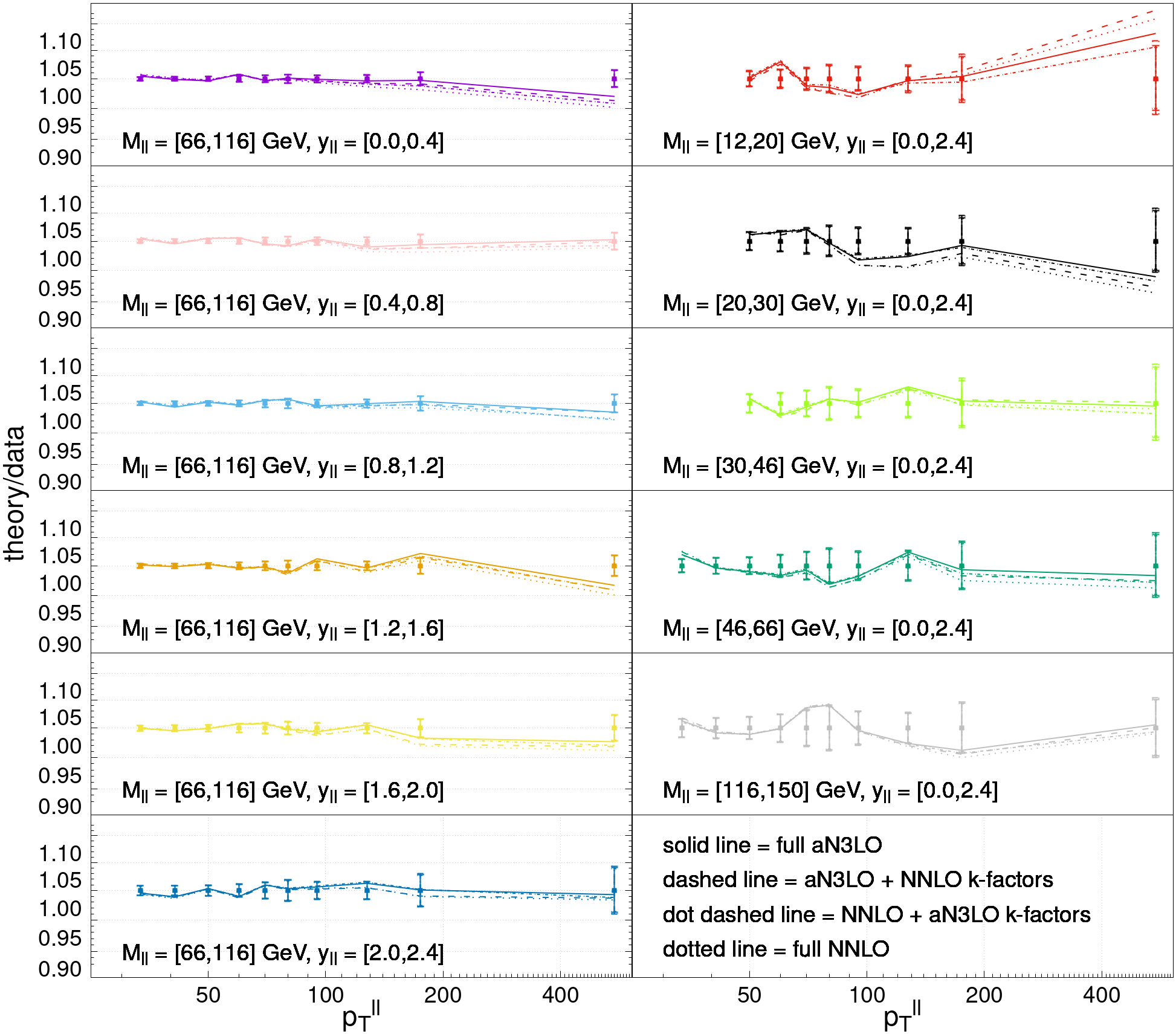} 
\caption{\sf Theory and data comparison at NNLO (dotted lines), NNLO with aN$^{3}$LO K-factors added (dot dashed lines), aN$^{3}$LO with NNLO K-factors (dashed lines) and aN$^{3}$LO (solid lines). The ratio of theory/data is shown and the points are the datapoints with their total uncorrelated error (quadrature sum of their statistical and uncorrelated systematic errors) only shown by the errorbars. On the left are the 6 double differential bins in $p_T^{ll}$ and $y_{ll}$ the $Z$ peak region, and on the right are the 5 single differential bins in $p_T^{ll}$ for the other mass bins. In this figure the data are after shifting by the fit  of the correlated systematic pulls.}
\label{fig:ZpT_datatheory_shifted}
\end{center}
\end{figure}

\newpage

\bibliography{references}{}

\begin{thebibliography}{10}

\bibitem{Bailey:2020ooq}
S.~Bailey, T.~Cridge, L.~A. Harland-Lang, A.~D. Martin, and R.~S. Thorne,
\newblock Eur. Phys. J. C {\bf 81}, 341 (2021), 2012.04684.

\bibitem{NNPDF:2021njg}
NNPDF, R.~D. Ball {\em et~al.},
\newblock Eur. Phys. J. C {\bf 82}, 428 (2022), 2109.02653.

\bibitem{Hou:2019efy}
T.-J. Hou {\em et~al.},
\newblock Phys. Rev. D {\bf 103}, 014013 (2021), 1912.10053.

\bibitem{ATLAS:2021vod}
ATLAS, G.~Aad {\em et~al.},
\newblock Eur. Phys. J. C {\bf 82}, 438 (2022), 2112.11266.

\bibitem{McGowan:2022nag}
J.~McGowan, T.~Cridge, L.~A. Harland-Lang, and R.~S. Thorne,
\newblock Eur. Phys. J. C {\bf 83}, 185 (2023), 2207.04739.

\bibitem{Cridge:2023ryv}
T.~Cridge, L.~A. Harland-Lang, and R.~S. Thorne,
\newblock (2023), 2312.07665.

\bibitem{Harland-Lang:2017ytb}
L.~A. Harland-Lang, A.~D. Martin, and R.~S. Thorne,
\newblock Eur. Phys. J. C {\bf 78}, 248 (2018), 1711.05757.

\bibitem{Bailey:2019yze}
S.~Bailey and L.~Harland-Lang,
\newblock Eur. Phys. J. C {\bf 80}, 60 (2020), 1909.10541.

\bibitem{Thorne:2022abv}
R.~Thorne, S.~Bailey, T.~Cridge, L.~Harland-Lang, and A.~D. Martin,
\newblock SciPost Phys. Proc. {\bf 8}, 018 (2022).

\bibitem{Jing:2023isu}
X.~Jing {\em et~al.},
\newblock Phys. Rev. D {\bf 108}, 034029 (2023), 2306.03918.

\bibitem{NNPDF:2017mvq}
NNPDF, R.~D. Ball {\em et~al.},
\newblock Eur. Phys. J. C {\bf 77}, 663 (2017), 1706.00428.

\bibitem{PDF4LHCWorkingGroup:2022cjn}
PDF4LHC Working Group, R.~D. Ball {\em et~al.},
\newblock J. Phys. G {\bf 49}, 080501 (2022), 2203.05506.

\bibitem{Cridge:2021qjj}
PDF4LHC21 combination group, T.~Cridge,
\newblock SciPost Phys. Proc. {\bf 8}, 101 (2022), 2108.09099.

\bibitem{Amoroso:2022eow}
S.~Amoroso {\em et~al.},
\newblock Acta Phys. Polon. B {\bf 53}, 12 (2022), 2203.13923.

\bibitem{ATLAS:2013jmu}
ATLAS, G.~Aad {\em et~al.},
\newblock JHEP {\bf 05}, 059 (2014), 1312.3524.

\bibitem{CMS:2012ftr}
CMS, S.~Chatrchyan {\em et~al.},
\newblock Phys. Rev. D {\bf 87}, 112002 (2013), 1212.6660,
\newblock [Erratum: Phys.Rev.D 87, 119902 (2013)].

\bibitem{CMS:2017jfq}
CMS, A.~M. Sirunyan {\em et~al.},
\newblock Eur. Phys. J. C {\bf 77}, 746 (2017), 1705.02628.

\bibitem{ATLAS:2017ble}
ATLAS, M.~Aaboud {\em et~al.},
\newblock JHEP {\bf 05}, 195 (2018), 1711.02692.

\bibitem{CMS:2022wdi}
CMS,
\newblock (2022), CMS-PAS-SMP-21-008.

\bibitem{Currie:2016bfm}
J.~Currie, E.~W.~N. Glover, and J.~Pires,
\newblock Phys. Rev. Lett. {\bf 118}, 072002 (2017), 1611.01460.

\bibitem{Gehrmann-DeRidder:2019ibf}
A.~Gehrmann-De~Ridder, T.~Gehrmann, E.~W.~N. Glover, A.~Huss, and J.~Pires,
\newblock Phys. Rev. Lett. {\bf 123}, 102001 (2019), 1905.09047.

\bibitem{AbdulKhalek:2020jut}
R.~Abdul~Khalek {\em et~al.},
\newblock Eur. Phys. J. C {\bf 80}, 797 (2020), 2005.11327.

\bibitem{Chen:2022tpk}
X.~Chen, T.~Gehrmann, E.~W.~N. Glover, A.~Huss, and J.~Mo,
\newblock JHEP {\bf 09}, 025 (2022), 2204.10173.

\bibitem{ATLAS:2015iiu}
ATLAS, G.~Aad {\em et~al.},
\newblock Eur. Phys. J. C {\bf 76}, 291 (2016), 1512.02192.

\bibitem{D0:2011jpq}
D0, V.~M. Abazov {\em et~al.},
\newblock Phys. Rev. D {\bf 85}, 052006 (2012), 1110.3771.

\bibitem{CDF:2007bvv}
CDF, A.~Abulencia {\em et~al.},
\newblock Phys. Rev. D {\bf 75}, 092006 (2007), hep-ex/0701051,
\newblock [Erratum: Phys.Rev.D 75, 119901 (2007)].

\bibitem{CMS:2015jdlc}
CMS, V.~Khachatryan {\em et~al.},
\newblock Eur. Phys. J. C {\bf 76}, 265 (2016), 1512.06212.

\bibitem{ATLAS:2014riz}
ATLAS, G.~Aad {\em et~al.},
\newblock JHEP {\bf 02}, 153 (2015), 1410.8857,
\newblock [Erratum: JHEP 09, 141 (2015)].

\bibitem{CMS:2014nvq}
CMS, S.~Chatrchyan {\em et~al.},
\newblock Phys. Rev. D {\bf 90}, 072006 (2014), 1406.0324.

\bibitem{CMS:2016lna}
CMS, V.~Khachatryan {\em et~al.},
\newblock JHEP {\bf 03}, 156 (2017), 1609.05331.

\bibitem{ATLAS:2017kux}
ATLAS, M.~Aaboud {\em et~al.},
\newblock JHEP {\bf 09}, 020 (2017), 1706.03192.

\bibitem{xfitterweb}
\texttt{https://xfitter.hepforge.org/data.html}.

\bibitem{Engin}
\texttt{Engin Eren and Katerina Lipka, private communication}.

\bibitem{Cacciari:2008gp}
M.~Cacciari, G.~P. Salam, and G.~Soyez,
\newblock JHEP {\bf 04}, 063 (2008), 0802.1189.

\bibitem{Sieber:2016kri}
G.~Sieber,
\newblock {\em {Measurement of Triple-Differential Dijet Cross Sections with
  the CMS Detector at 8 TeV and PDF Constraints}},
\newblock PhD thesis, KIT, Karlsruhe, 2016.

\bibitem{Klaus}
\texttt{Klaus Rabbertz, private communication}.

\bibitem{Carli:2010rw}
T.~Carli {\em et~al.},
\newblock Eur. Phys. J. C {\bf 66}, 503 (2010), 0911.2985.

\bibitem{Kluge:2006xs}
T.~Kluge, K.~Rabbertz, and M.~Wobisch,
\newblock {FastNLO: Fast pQCD calculations for PDF fits},
\newblock in {\em {14th International Workshop on Deep Inelastic Scattering}},
  pp. 483--486, 2006, hep-ph/0609285.

\bibitem{Britzger:2012bs}
fastNLO, D.~Britzger, K.~Rabbertz, F.~Stober, and M.~Wobisch,
\newblock {New features in version 2 of the fastNLO project},
\newblock in {\em {20th International Workshop on Deep-Inelastic Scattering and
  Related Subjects}}, pp. 217--221, 2012, 1208.3641.

\bibitem{Currie:2018xkj}
J.~Currie {\em et~al.},
\newblock JHEP {\bf 10}, 155 (2018), 1807.03692.

\bibitem{Dittmaier:2012kx}
S.~Dittmaier, A.~Huss, and C.~Speckner,
\newblock JHEP {\bf 11}, 095 (2012), 1210.0438.

\bibitem{Cridge:2021pxm}
T.~Cridge, L.~A. Harland-Lang, A.~D. Martin, and R.~S. Thorne,
\newblock Eur. Phys. J. C {\bf 82}, 90 (2022), 2111.05357.

\bibitem{CMS:2015jdl}
CMS, V.~Khachatryan {\em et~al.},
\newblock Eur. Phys. J. C {\bf 76}, 265 (2016), 1512.06212.

\bibitem{CMS:2015rld}
CMS, V.~Khachatryan {\em et~al.},
\newblock Eur. Phys. J. C {\bf 75}, 542 (2015), 1505.04480.

\bibitem{ATLAS:2015lsn}
ATLAS, G.~Aad {\em et~al.},
\newblock Eur. Phys. J. C {\bf 76}, 538 (2016), 1511.04716.

\bibitem{ATLAS:2016pal}
ATLAS, M.~Aaboud {\em et~al.},
\newblock Phys. Rev. D {\bf 94}, 092003 (2016), 1607.07281,
\newblock [Addendum: Phys.Rev.D 101, 119901 (2020)].

\bibitem{CMS:2017iqf}
CMS, A.~M. Sirunyan {\em et~al.},
\newblock Eur. Phys. J. C {\bf 77}, 459 (2017), 1703.01630.

\bibitem{ATLAS:2016nqi}
ATLAS, M.~Aaboud {\em et~al.},
\newblock Eur. Phys. J. C {\bf 77}, 367 (2017), 1612.03016.

\bibitem{Chen:2021vtu}
X.~Chen {\em et~al.},
\newblock Phys. Rev. Lett. {\bf 128}, 052001 (2022), 2107.09085.

\bibitem{Boughezal:2017nla}
R.~Boughezal, A.~Guffanti, F.~Petriello, and M.~Ubiali,
\newblock JHEP {\bf 07}, 130 (2017), 1705.00343.

\bibitem{Campbell:2002tg}
J.~M. Campbell and R.~Ellis,
\newblock Phys. Rev. D {\bf 65}, 113007 (2002), hep-ph/0202176.

\bibitem{Gehrmann-DeRidder:2016cdi}
A.~Gehrmann-De~Ridder, T.~Gehrmann, E.~W.~N. Glover, A.~Huss, and T.~A. Morgan,
\newblock JHEP {\bf 07}, 133 (2016), 1605.04295.

\bibitem{Bizon:2018foh}
W.~Bizo\'n {\em et~al.},
\newblock JHEP {\bf 12}, 132 (2018), 1805.05916.

\bibitem{NMC}
NMC, M.~Arneodo {\em et~al.},
\newblock Nucl. Phys. {\bf B483}, 3 (1997), hep-ph/9610231.

\bibitem{NuTev}
NuTeV, M.~Tzanov {\em et~al.},
\newblock Phys. Rev. {\bf D74}, 012008 (2006), hep-ex/0509010.

\bibitem{E866DYrat}
NuSea, R.~S. Towell {\em et~al.},
\newblock Phys. Rev. {\bf D64}, 052002 (2001), hep-ex/0103030.

\bibitem{Dimuon}
NuTeV, M.~Goncharov {\em et~al.},
\newblock Phys. Rev. {\bf D64}, 112006 (2001), hep-ex/0102049.

\bibitem{H1+ZEUS}
H1, ZEUS, F.~Aaron {\em et~al.},
\newblock JHEP {\bf 01}, 109 (2010), 0911.0884.

\bibitem{D0Wnumu}
D0, V.~M. Abazov {\em et~al.},
\newblock Phys. Rev. {\bf D88}, 091102 (2013), 1309.2591.

\bibitem{LHCb-Zee}
LHCb, R.~Aaij {\em et~al.},
\newblock JHEP {\bf 02}, 106 (2013), 1212.4620.

\bibitem{LHCb-WZ}
LHCb, R.~Aaij {\em et~al.},
\newblock JHEP {\bf 06}, 058 (2012), 1204.1620.

\bibitem{ATLAShighmass}
ATLAS, G.~Aad {\em et~al.},
\newblock Phys. Lett. {\bf B725}, 223 (2013), 1305.4192.

\bibitem{LHCbZ8}
LHCb, R.~Aaij {\em et~al.},
\newblock JHEP {\bf 05}, 109 (2015), 1503.00963.

\bibitem{CMS7Wpc}
CMS, S.~Chatrchyan {\em et~al.},
\newblock JHEP {\bf 02}, 013 (2014), 1310.1138.

\bibitem{ATLASWZ7f}
ATLAS, M.~Aaboud {\em et~al.},
\newblock Eur. Phys. J. C {\bf 77}, 367 (2017), 1612.03016.

\bibitem{D0Wasym}
D0, V.~M. Abazov {\em et~al.},
\newblock Phys. Rev. Lett. {\bf 112}, 151803 (2014), 1312.2895,
\newblock [Erratum: Phys.Rev.Lett. 114, 049901 (2015)].

\bibitem{ATLASsdtop}
ATLAS, G.~Aad {\em et~al.},
\newblock Eur. Phys. J. C {\bf 76}, 538 (2016), 1511.04716.

\bibitem{ATLASWjet}
ATLAS, M.~Aaboud {\em et~al.},
\newblock JHEP {\bf 05}, 077 (2018), 1711.03296.

\bibitem{CMS8ttDD}
CMS, A.~M. Sirunyan {\em et~al.},
\newblock Eur. Phys. J. C {\bf 77}, 459 (2017), 1703.01630.

\bibitem{ATLASW8}
ATLAS, G.~Aad {\em et~al.},
\newblock Eur. Phys. J. C {\bf 79}, 760 (2019), 1904.05631.

\bibitem{CMS276jets}
CMS, V.~Khachatryan {\em et~al.},
\newblock Eur. Phys. J. C {\bf 76}, 265 (2016), 1512.06212.

\bibitem{CMSttbar08_ytt}
CMS, V.~Khachatryan {\em et~al.},
\newblock Eur. Phys. J. C {\bf 75}, 542 (2015), 1505.04480.

\bibitem{ATLAS8Z3D}
ATLAS, M.~Aaboud {\em et~al.},
\newblock JHEP {\bf 12}, 059 (2017), 1710.05167.

\end{thebibliography}
\bibliographystyle{h-physrev}

\end{document}